\documentclass[useAMS,usenatbib]{mn2e}
\usepackage{graphicx}
\usepackage{graphicx,xspace}
\usepackage{epsf}
\usepackage{amsmath}
\usepackage{color}
\usepackage{epsfig,epstopdf}

\def\gsim { \lower .75ex \hbox{$\sim$} \llap{\raise .27ex \hbox{$>$}} }
\def\lsim { \lower .75ex \hbox{$\sim$} \llap{\raise .27ex \hbox{$<$}} }

\newcommand{\FOF}{FoF\xspace}
\newcommand{\FoF}{FoF\xspace}
\newcommand{\Dhalo}{Dhalo\xspace}
\newcommand{\Dhaloes}{Dhaloes\xspace}
\newcommand{\Dhalos}{Dhaloes\xspace}
\newcommand{\DinF}{\Dhalo in \FOF halo\xspace}
\newcommand{\FinD}{\FOF in \Dhalo}
\newcommand{\subhalo}{subhalo\xspace}
\newcommand{\subhaloes}{subhaloes\xspace}
\newcommand{\subgroup}{subhalo\xspace}
\newcommand{\subgroups}{subhaloes\xspace}
\newcommand{\Subgroup}{Subhalo\xspace}
\newcommand{\Subgroups}{Subhaloes\xspace}
\newcommand{\Subhaloes}{Subhaloes\xspace}
\newcommand{\SUBFIND}{\textsc{subfind}\xspace}
\newcommand{\subfind}{\textsc{subfind}\xspace}
\newcommand{\GALFORM}{\textsc{galform}\xspace}
\newcommand{\galform}{\textsc{galform}\xspace} 

\newcommand{\Msol}{$h^{-1}$M$_{\odot}$\xspace}
\renewcommand\tabcolsep{0.2cm}

\title[Hierarchial N-body Dark Matter Haloes]{N-body Dark Matter
  Haloes with simple Hierarchical Histories}
\author[Jiang et al.] {\parbox{18cm}{
Lilian Jiang, John C.~Helly, Shaun Cole, Carlos S.~Frenk
}\vspace{0.3cm}\\
{Institute for Computational Cosmology, Department of Physics, University of
  Durham, South Road, Durham  DH1 3LE, UK}
  }

\begin{document}
\pagerange{\pageref{firstpage}--\pageref{lastpage}} \pubyear{2014}
\maketitle
\label{firstpage}


\begin{abstract} 
We present a new algorithm which groups the subhaloes found 
in cosmological N-body simulations by
structure finders such as \SUBFIND\ into dark matter haloes 
whose formation histories are strictly hierarchical.
One advantage of these `\Dhaloes' over the
commonly used friends-of-friends (\FoF) haloes is that they retain
their individual identity in cases when \FoF haloes are artificially
merged by tenuous bridges of particles or by an overlap of their outer
diffuse haloes. \Dhaloes are thus well-suited for modelling galaxy
formation and their merger trees form the basis of the Durham
semi-analytic galaxy formation model, \GALFORM. Applying the \Dhalo
construction to the $\Lambda$CDM Millennium-2 simulation we find that
approximately 90\% of \Dhaloes have a one-to-one, bijective match with a
corresponding \FoF halo. The remaining 10\% are typically secondary
components of large \FoF haloes. Although the mass functions of both
types of haloes are similar, the mass of \Dhalos correlates much more
tightly with the virial mass, $M_{200}$, than \FoF haloes.
Approximately 80\% of \FoF and bijective and non-bijective \Dhaloes are
relaxed according to standard criteria.  For these relaxed haloes all
three types have similar concentration--$M_{200}$ relations and, at
fixed mass, the concentration distributions are described accurately
by log-normal distributions. 
\end{abstract}

\begin{keywords}
methods: numerical - galaxies: haloes - cosmology: theory - dark matter.
\end{keywords}

\section{INTRODUCTION}

In hierarchical dark matter dominated cosmologies, such as standard
$\Lambda$CDM, galaxy formation is believed to be intimately linked to
the formation and evolution of dark matter haloes. Baryonic gas falls
into dark matter haloes, cools and settles into centrifugally
supported star forming discs \citep{binney77, rees77,
white78,white91,kw,cole94,somerville99,benson03}.
 Thus the evolution of the galaxy population is
driven by the evolution of the population of dark matter haloes which
grow hierarchically via mergers and accretion.  Thus to model galaxy
formation one must first have an accurate model of the evolution of
dark matter haloes.

The formation and evolution of dark matter haloes from cosmological
initial conditions in large representative volumes can now be
routinely and reliably simulated using a variety of N-body codes
\citep[e.g.][]{springel05a}. In contrast, simulations of the evolution
of the baryonic component are much more uncertain with gross
properties depending on the details of uncertain sub-grid physics as
well as on the limitations of numerical hydrodynamics 
\citep{schaye10,creasey11}.
 Hence a useful and complementary approach is
semi-analytic galaxy formation \citep[e.g.][]{white91,cole91,lacey91,kw,
cole94,cole00, somerville99,somerville08,bower06,benson10}
in which one starts with the
framework provided by the dark matter halo evolution and uses analytic
models to follow the processes of galaxy formation that occur within
these haloes. The key starting point for this approach is halo merger
trees which quantify the hierarchical growth of individual dark matter
haloes.

In $\Lambda$CDM the first self-bound objects to form are haloes
with masses of around an Earth mass corresponding to the small scale
thermal cut off in the CDM power spectrum \citep{green04}. In a
cosmological N-body simulation the mass scale of the first generation
of haloes is instead set by the mass resolution of the
simulation. Subsequent generations of haloes form by mergers of
earlier generations of haloes plus some smoothly accreted
material. The merging process does not produce a completely relaxed
smooth halo and the remnants of the earlier generation of haloes are
often detectable as self-bound substructures (\subhaloes) within the
new halo.  Thus it is important to distinguish between haloes and the
\subhaloes that they contain which are the remnants of early
generations of now merged haloes.  A variety of algorithms which can
identify these \subhaloes in N-body simulations has been devised
\citep{onions12}. These substructure finders are capable of
detecting arbitrary levels of nested \subhaloes within \subhaloes and
in most cases also identify the background mass distribution in a halo
as a subhalo. In this work we refer to all of the groups
identified by such substructure finders as ``\subhaloes'' and merger
trees constructed by identifying a descendant for each \subhalo as
``subhalo merger trees''. \cite{srisawat13} compare a range of methods
for the production of subhalo merger trees. The algorithm we use to
determine subhalo descendants in this paper is included in the
comparison under the name {\sc{D-Trees}}. 

 To construct the halo merger trees
needed by semi-analytic models it is not sufficient to just track \subhaloes
between simulation outputs (e.g. by tracking their consituent
particles), one also needs to identify their host haloes.  For
instance when a galaxy cluster forms it is normally assumed that while
the galaxies remain in their individual \subhaloes the diffuse gas
surrounding them and gas blown out of the galaxies by SN driven winds
is not retained by the individual \subhaloes but instead joins the
common intra-cluster medium of the surrounding halo of the galaxy
cluster. Another issue that has to be addressed when building merger
trees for use in galaxy formation models is that structure formation
for the collisionless material in N-body simulations is not strictly
hierarchical. Hence occasionally when two haloes merge the subhalo
resulting from the smaller progenitor can pass straight through the
main halo and escape to beyond its virial radius.
For the galaxy formation process to be followed it is necessary to
retain the association between these two separated subhaloes so that an
appropriate physical model can be applied to their 
diffuse collisional gas which would not have separated after the
merger. Merger trees that are useful for galaxy formation modelling
have to take account of these considerations \citep{knebe13}. 
The \Dhalo algorithm
which we present produces a set of haloes which is strictly
hierarchical in the sense that once a subhalo becomes a component of
a \Dhalo it never subsequently demerges.

It is now quite common for semi-analytic models to use halo merger
trees extracted directly from N-body simulations
\citep{springel01,helly03,hatton03,bower06,munoz09,koposov09,busha10,
  maccio10,guo11}. 
There are many choices to be made both in defining the
halo catalogues and in constructing the links between haloes at
different times. \cite{knebe11} and \cite{knebe13} have found    
significant differences in even the most basic properties (e.g the
halo mass function) of halo catalogues constructed with different  
group finding codes. Additionally, these halo 
catalogues can often be modified by the 
procedure of constructing the merger trees as some of the algorithms
break up or merge haloes together in order to achieve a more 
consistent membership over time \citep{helly03,behroozi13}.
So, for example,
even if one starts with standard Friends-of-Friends (\FoF ) groups
\citep{davis85} the process of building the merger trees can alter
the abundance and properties of the haloes.

Semi-analytic models such as \GALFORM have the option of using
information extracted directly from an N-body simulation or using
Monte Carlo methods 
\citep[see][for a comparison of different algorithms]{jiangvdb13}
which make use of statistical descriptions of
N-body results such as analytic halo mass functions
\citep[e.g.][]{sheth99,jenkins01,evrard02,white01b,
reed03, linder03, lokas04, warren06,heitmann06,reed07,
lukic09,tinker08,Boylan-Kolchin09, crocce10, courtin10,
bhattacharya11,watson13}  and models for the
distribution of the concentrations of halo mass profiles
\citep[e.g.][]{nfw95,nfw96,bullock01,eke01,maccio08}.
These statistical descriptions are often based on the abundance and properties of \FoF
haloes and so may not be directly applicable to the catalogues of
haloes that result from the application of a specific merger tree
algorithm. The internal structure of the dark matter haloes strongly
influences galaxy formation models. Often the gas density profiles
within dark matter haloes are assumed to be related to the dark matter
profile, e.g. through hydrostatic equilibrium and these influence the
rate at which gas cools onto the central galaxy. In addition the
central potential of the dark matter halo effects the size and
circular velocity of the central galaxy which in turn can have a
strong effect on the expulsion of gas from the galaxy via SN
feedback.   
Hence for semi-analytic galaxy formation modelling it is
important to adopt models of the individual haloes that are consistent
with the haloes that appear in the merger trees used by semi-analytic model.

In this paper we present a detailed description of the latest N-body
merger tree algorithm that has been developed for use with the
semi-analytic code \GALFORM. The algorithm is an improvement over the
earlier version, described in \cite{merson12}, which was run on the
Millennium simulation \citep{springel05a} and widely exploited in a
range of applications \citep{bower06,font08,kim11,merson12}.  The
resulting differences between the two algorithms are very small when applied to
relatively low resolution simulations such as the Millennium, but the
improvements in the new algorithm do a better job of tracking halo
descendants in high resolution simulations such as the Millennium II
\citep{Boylan-Kolchin09} and Aquarius simulations \citep{springel08}.
The starting point for our merger trees are \FoF haloes that are
decomposed into \subhaloes, distinct self-bound structures, by the
substructure finder, \SUBFIND \citep{springel01}.  \Subhaloes are
tracked between output times and agglomerated into a new set of haloes,
dubbed \Dhaloes, that have consistent membership over time in the sense
that once a \subhalo is accreted by a \Dhalo it never demerges. In this
process we also split some \FoF haloes into two or more \Dhaloes when
\SUBFIND substructures are well separated and only linked into a single
\FoF halo by bridges of low density material.

\begin{figure}
\includegraphics[width=8.5cm]{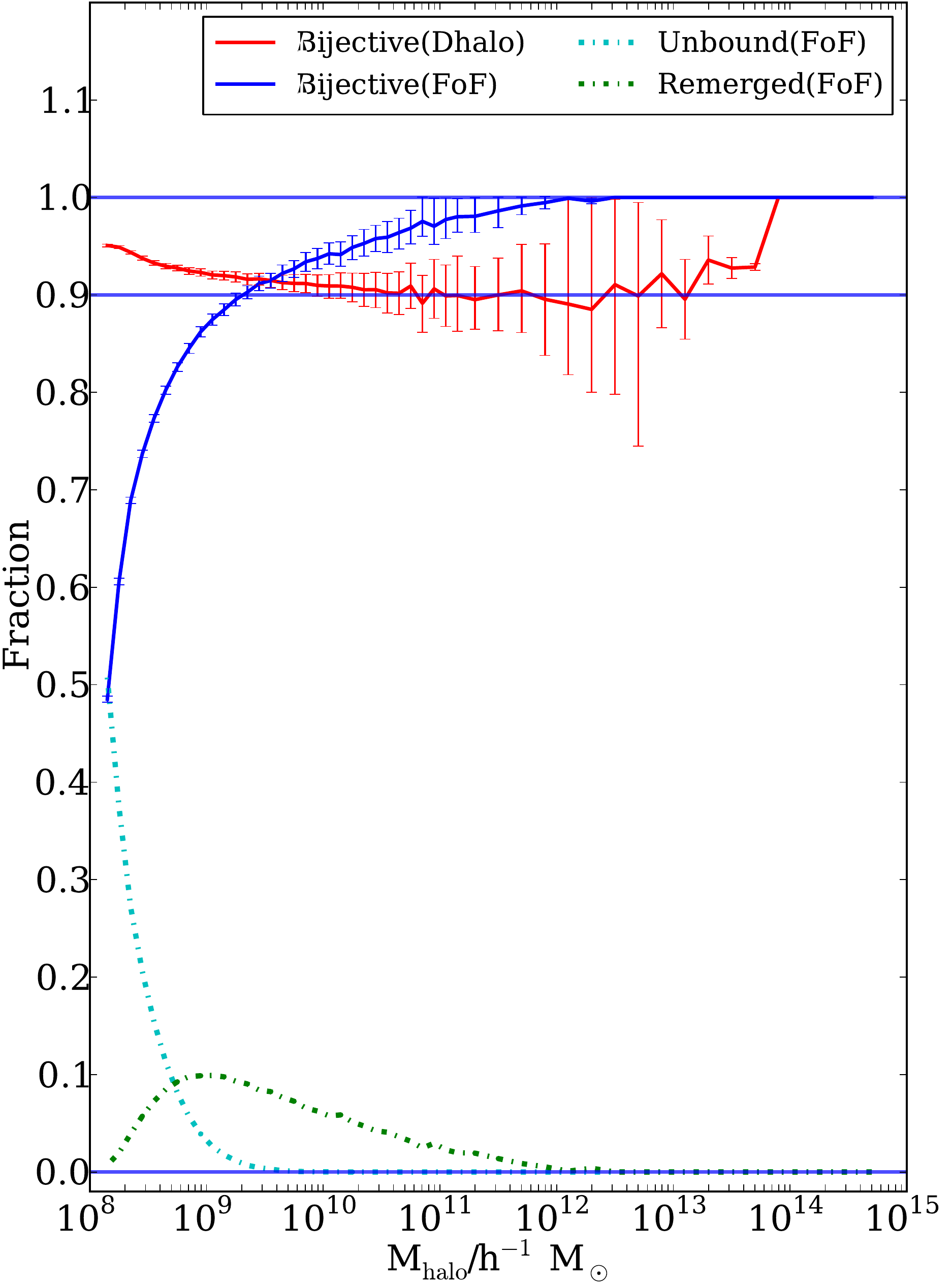}%
\caption{The upper two curves, with bootstrap error bars, show the
fraction of \Dhalo (red) and \FOF haloes (blue) in the MSII catalogues
that have a bijective (a unique one-to-one) match as a function of
their respective \Dhalo or \FoF halo mass. The lower two curves show
the fraction of \FOF haloes that do not contain a self-bound
substructure (cyan) and the fraction whose main \subhaloes are
remerged by the \Dhalo algorithm to form part of a more massive \Dhalo
(green).}
\label{fig:f1} 
\end{figure}

Our paper is structured as follows.  In Section~\ref{sec:Haloes} we
briefly outline the new merger tree algorithm (full details are given
in Appendix~\ref{app:Trees}) and its application to the Millennium~II
simulation.  In Section~\ref{sec:single} we compare and contrast the
properties of the resulting \Dhaloes with the more commonly used \FOF
haloes \citep{davis85}.  We show specific rare examples where \Dhaloes
and their matched \FOF counterparts exhibit gross differences either
one \FOF halo being decomposed into several \Dhaloes or vice versa. We
also examine the distribution of mass ratios for matching \Dhalo and
\FoF pairs.  Then in Section~\ref{sec:stat} we compare statistical
properties of halo populations including halo mass functions and their
concentration--mass relation. We conclude in Section~\ref{sec:conc}.

\section{Halo Catalogues}
\label{sec:Haloes}

Immediately below, Section~\ref{sec:MillII}, we summarise the
specification of the Millennium~II simulation which we use to test
and illustrate the application of our merger tree algorithm. We then
give a brief outline of the construction of the merger trees 
and their constituent haloes with the
complete specification detailed in Appendix ~\ref{app:Trees}.

\subsection{The Millennium-II Simulation}
\label{sec:MillII}

The Millennium-II (MSII) simulation\footnote{
The Millennium-II simulation data will be available from an SQL relational database 
that can be accessed at \qquad \qquad \qquad \qquad
http://galaxy-catalogue.dur.ac.uk:8080/Millennium .} 
\citep{Boylan-Kolchin09} was
carried out with the {\sc gadget3} N-body code, which uses a ``TreePM"
method to calculate gravitational forces.  The MSII is a cosmological
simulation of the standard $\Lambda$CDM cosmology in a periodic box of
side $L_{\rm box}$= 100$h^{-1}$ Mpc containing $N = 2160^{3}$
particles of mass $6.95 \times 10^6 $~\Msol.  
The cosmological parameters for the MSII are: $\Omega_{\rm
m} = 0.25$, $\Omega_{\rm b} = 0.045$, $h = 0.73$, $\Omega_{\Lambda} =
0.75$, $n = 1$ and $\sigma_{8} = 0.9$. Here $\Omega_{\rm m}$ denotes
the total matter density in units of the critical density, $\rho_{\rm
crit} = 3H_{0}^{2}/(8\pi G)$. $\Omega_{\rm b}$ and $\Omega_{\Lambda}$
denote the densities of baryons and dark energy at the present
day in units of the critical density.
 The Hubble constant is $H_{0} = 100 h\, $km$\,$s$^{-1}$~Mpc$^{-1}$, $n$
is the primordial spectral index and $\sigma_{8}$ is the rms density
fluctuation within a sphere of radius $8 h^{-1}$Mpc extrapolated
to $z$ = 0 using linear theory.  
These cosmological parameters are consistent with a
combined analysis of the 2dFGRS~\citep[]{colless01, percival01} and
first year WMAP data~\citep[]{spergel03, sanchez06}.

\subsection{Building Merger Trees}
\label{sec:Trees}

The first step in building our merger trees is the construction of
catalogues of both \FoF haloes \citep{davis85} and their internal
self-bound substructures\footnote{Here we identify the \subhaloes
using the \SUBFIND\ algorithm (Springel et al 2001). However this is
not the only option and there is now a large literature
\citep[see][]{onions12} on alternative methods of identifying
self-bound structures. Some of these are highly sophisticated and use
full 6D phase space information to disentangle spatial coincident
subhaloes \citep{diemand06,behroozi12}. As an example, we have
experimented with building \Dhaloes by applying the \Dhalo algorithm
from Appendix~A3 onwards but with \subfind \subhalo merger trees
replaced by those defined by the Hierarchical Bound Tracing (HBT)
algorithm of \citep{han12}.  We find that the properties of the \Dhalo
merger trees and the galaxies that result after they are processed by
\GALFORM are extremely similar.}, \subhaloes, as identified by
\SUBFIND \citep{springel01}.  The second step is to build \SUBFIND
merger trees by tracking particles between output snapshots to
determine the descendant of each \subhalo. Occasionally \SUBFIND fails
to find a substructure as it transits through the core of a larger
halo.  To avoid this resulting in the premature merging of
substructures we have developed an algorithm
(Appendix~\ref{app:build}) that looks several snapshots ahead to
robustly link progenitor and descendant \subhaloes.  A similar
approach was adopted by \cite{behroozi13} to construct self-consistent
merger trees for the Bolshoi simulations \citep{klypin11}.  The third
step is to partition these \SUBFIND merger trees into discrete
branches. A new branch begins whenever a new \subhalo forms and
continues for as long as the \subhalo exists in the simulation.  When
a merger occurs we decide which of the progenitor \subhaloes survives
the merger by determining which progenitor contributed the most bound
part of the descendant (see Appendix~\ref{app:findingdescendants}
). The branch corresponding to this progenitor continues, while the
other progenitor's branch ends.  The final step is to bundle these
branches together to define the composite \Dhaloes and their merger
trees. Here our algorithm (described in full in
Appendix~\ref{app:Dhalo}) defines collections of \subhaloes embedded
hierarchically within each other as a single \Dhalo, but excludes
neighbouring \subhaloes that may be part of the same \FOF group, but
are only linked in by a bridge of low density material or \subhaloes
that are beginning the process of merging but have not yet lost a
significant amount of mass.  Subhaloes are grouped into \Dhaloes in
such a way that once a subhalo becomes part of a \Dhalo it remains a
component of that \Dhalo's descendants at all later times at which the
subhalo survives, even if it is a satellite component that takes it
temporarily outside the corresponding \FoF halo.  All of a Dhalo's
\subhaloes which survive at a later snapshot must belong to the same
\Dhalo at that snapshot. We take this to be the descendant of the
\Dhalo. This defines the \Dhalo merger trees.  The mass of a \Dhalo is
simply the sum of the masses of its component \subhaloes.\footnote{As
a subhalo can, by definition, only belong to one \Dhalo and as
particles can only belong to one \SUBFIND \subhalo this means that
\Dhalos are exclusive in the sense that no particles can belong to
more than one \Dhalo.}


\begin{figure*}

\begin{center}

\begin{tabular}{cc}
\resizebox{8cm}{!}{\includegraphics{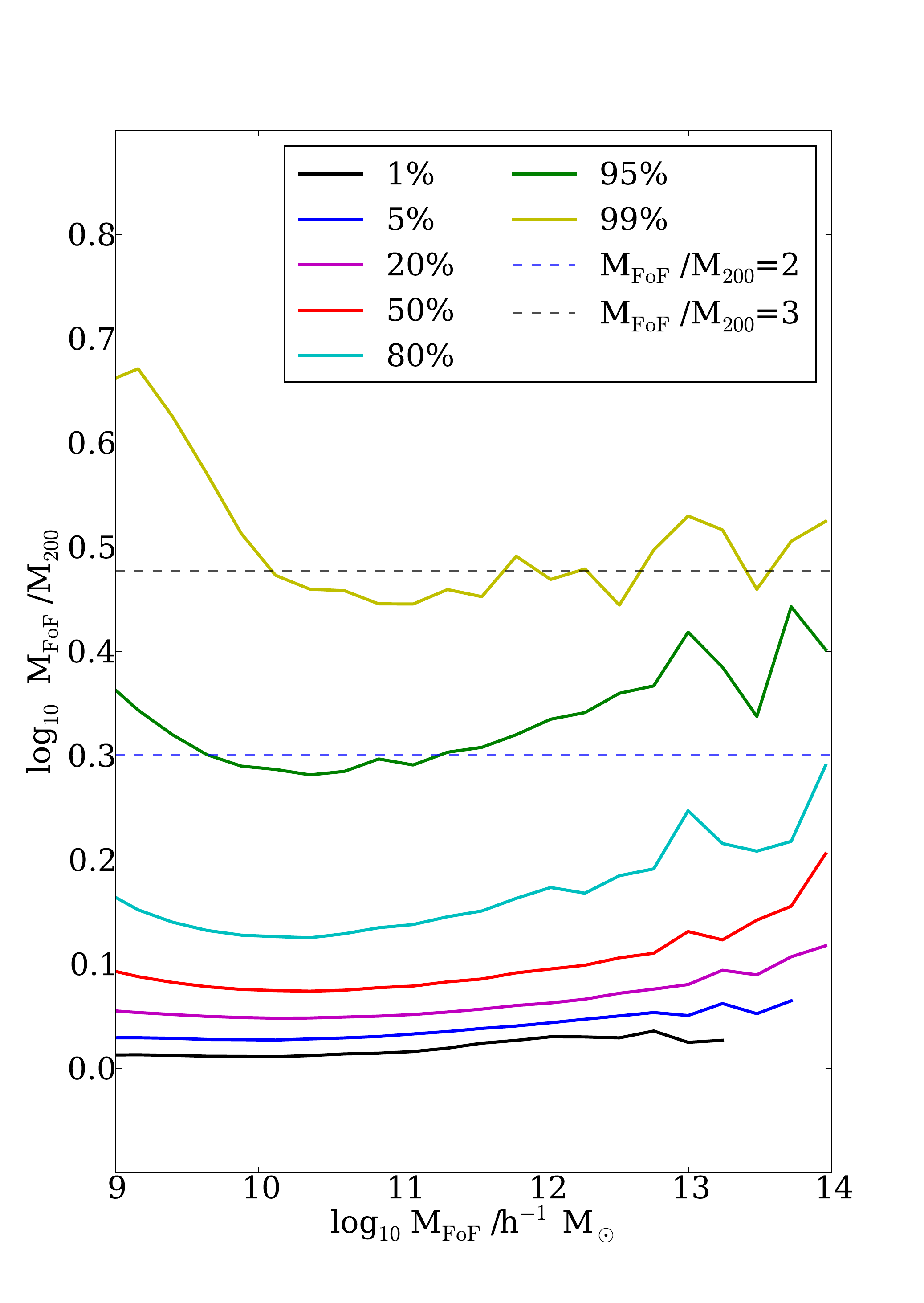}}%
\resizebox{8cm}{!}{\includegraphics{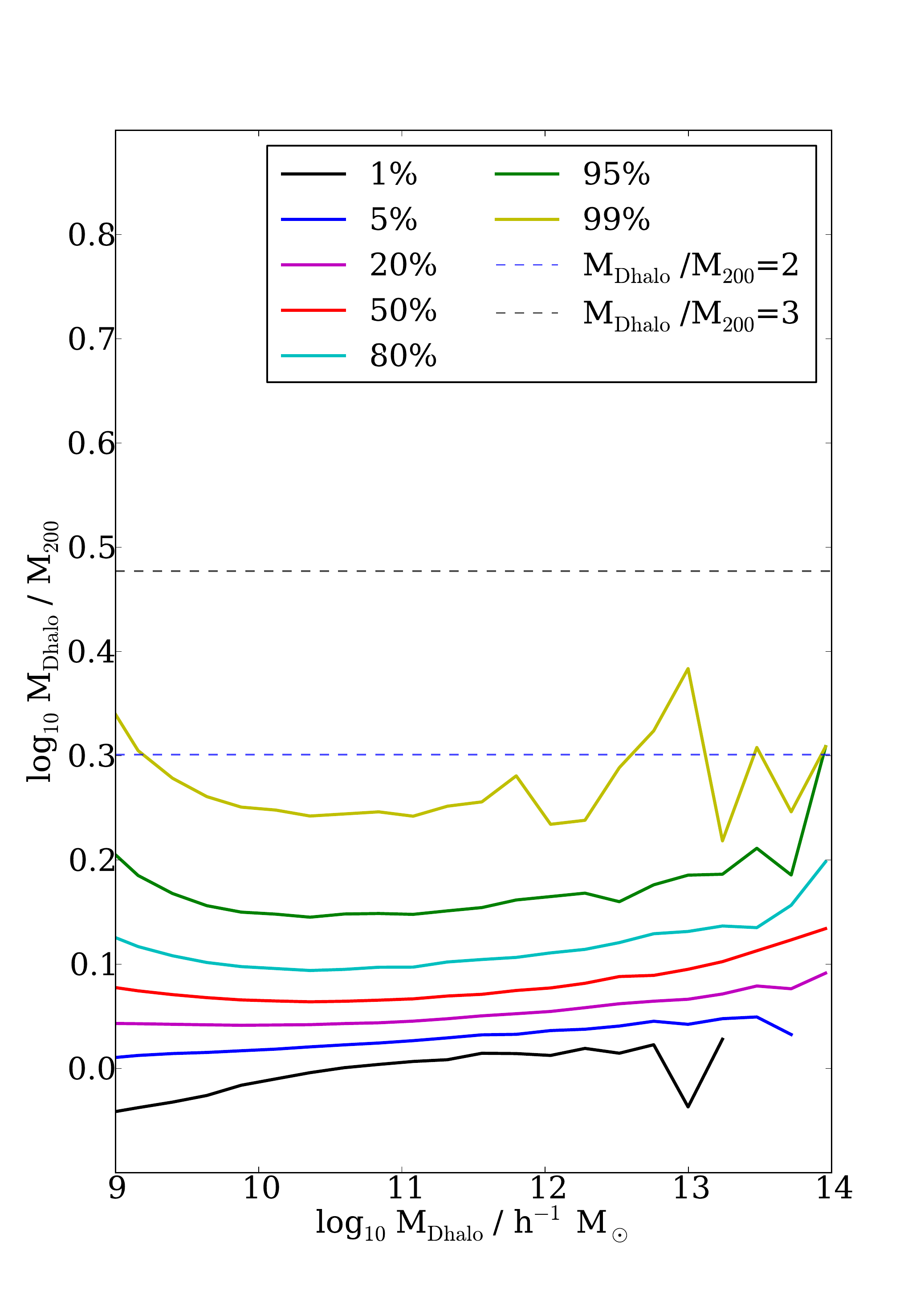}}
\end{tabular}
\end{center}
\caption{The left panel shows the median, 1, 5, 20, 80, 95, 99
  percentiles of the distribution of the mass ratios between \FoF halo
  mass, M$_{\rm FoF}$ and virial mass, M$_{200}$ as a function of \FoF
  halo mass for haloes identified using the \FoF group finder.  The
  right panel shows the same percentiles for the distribution of the
  mass ratio between \Dhalo mass, M$_{\rm Dhalo}$ and virial mass,
  M$_{\rm 200}$, as a function of \Dhalo mass for haloes identified
  using the \Dhalo group finder.  The blue dashed line in both panels
  shows where M$_{\rm Halo}$/M$_{\rm 200}$ =2.0 and the black one
  M$_{\rm Halo}$/M$_{\rm 200}$=3.0.}
\label{fig:ms2}
\end{figure*}

\section{Comparison of \FOF and \Dhaloes}
\label{sec:single}

\subsection{Bijectively matched \FOF and \Dhaloes}

The properties of \FOF haloes, especially those defined by the
conventional linking length parameter of $b=0.2$ (the linking length
is defined as $b$ times the mean inter-particle separation), are well
documented in the literature \citep[e.g.][]{frenk88, lacey94, summers95, audit98, 
huchra82, press82, einasto84, davis85, frenk88, 
lacey94, klypin99, jenkins01, warren06, eke04,gott07} and such
haloes are widely used as the starting point for relating the dark
matter and galaxy distributions\citep{peacock00,seljak00,berlind02}.
Thus as the semi-analytic
model {\sc galform} \citep{bower06,font08,font11,lagos11} instead uses \Dhaloes as its
starting point, it is interesting to contrast the properties of 
haloes defined by these two algorithms. 

As described in Section~\ref{sec:Haloes}, \FoF haloes are decomposed
by \subfind into \subhaloes and those are then regrouped into
\Dhaloes. Hence for every \FoF halo, we can find its matching \Dhalo
by finding which \Dhalo contains the most massive \subhalo from the
\FoF group. We can perform this matching the other way round by
finding the \FoF halo containing the most massive \subhalo from the
\Dhalo. In cases where the most massive \subhalo of a \FoF halo is
also the most massive \subhalo of a \Dhalo, these two matching
procedures produce identical associations. We refer to such cases as
bijective matches.

Before comparing the properties of this subset of bijectively matched
Dhaloes and \FOF haloes we first quantify how representative they are
by looking at the fraction of each set of haloes that have these
bijective matches.  The two upper curves in Fig.~\ref{fig:f1} show 
the dependence of the
bijective fraction of \Dhaloes on \Dhalo mass and \FoF haloes on \FoF
mass. The first thing to note is that the fraction of bijectively
matched \Dhaloes is large, being 90\% or greater over the full range
from $10^8$ to $10^{14}$~\Msol and so to a first approximation there is
a good correspondence between \FOF and \Dhaloes. Above $3 \times
10^{10}$~\Msol about 10\% of the \Dhaloes do not have a bijective
match which means they instead represent secondary fragments of more
massive \FOF haloes that the \Dhalo algorithm has split into two or
more \subhaloes. Below $3 \times 10^{10}$~\Msol this non-bijective
fraction drops indicating that lower mass \FOF haloes are less likely
to be split into two or more comparable mass \Dhaloes. This behaviour
is consistent with the results of \cite{lukic09} who found that
15-20\% of \FoF haloes are irregular structures that have two or more
major components linked together by low density bridges and that this
fraction is an increasing function of halo mass. This is also to be
expected in the hierarchical merging picture as the most massive
haloes formed most recently and so are the least dynamically relaxed.

For the \FoF haloes with mass above $10^{12}$~\Msol the bijectively
matched fraction is unity, indicating that the most massive \subhalo
of such \FoF haloes together with the \subhaloes embedded within it
always gives rise to a \Dhalo.  Below $10^{12}$~\Msol the the
bijective fraction begins to decrease steadily with decreasing
mass. This happens because as the \FOF mass decreases there is an
increasing probability that the progenitor of this \FOF halo has
previously passed through a more massive neigbouring halo and this
results in the Dhalo algorithm remerging the \FOF halo with its more
massive neighbour. This fraction of \FOF haloes that are remerged to
form part of a more massive \Dhalo is shown by the green curve in
Fig.~\ref{fig:f1}. As one approaches 10$^{8}$~\Msol ($\sim$$15$
particles) the bijective
fraction plummets as at very low masses many of the \FOF haloes are
not self-bound and so do not contain any \subhaloes from which to build
a \Dhalo . The fraction of \FOF haloes which do not contain a
self-bound substructure is shown by the cyan curve in
Fig.~\ref{fig:f1} and can be seen to reach 50\% at at a \FOF mass
of 20 particles.

\subsubsection{Virial Masses}

It is conventional to define the virial mass, $M_{\rm vir}$, and
associated virial radius, $r_{\rm vir}$, of a dark matter halo using a
simple spherical overdensity criterion centred on the potential minimum. 
\begin{equation}
M_{\rm vir} =\frac{4}{3}\pi \Delta \, \rho_{\rm crit} \,
r_{\rm vir}^3
\label{eq:rvir}
\end{equation}
where $\rho_{\rm crit}$ is the cosmological critical density and
$\Delta$ is the specified overdensity. In applying this definition we
adopt $\Delta=200$ and include all the particles inside this spherical
volume, not only the particles grouped by the \FOF or \Dhalo
algorithm, to define the enclosed mass, $M_{200}$
, and associated radius
$r_{200}$. This choice is largely a matter of convention but has been
shown to roughly correspond to boundary at which the haloes are 
in approximate dynamical equilibrium \cite[e.g.][]{cole96}.

If the halo finding algorithm has succeeded in partitioning the dark
matter distribution into virialized haloes we would expect to see a good
correspondence between the grouped mass of the halo and $M_{200}$.
For instance, as \FoF haloes are essentially bounded by an isodensity
contour, whose value is set by the linking parameter \citep{davis85}, 
then if they have relaxed quasi-spherical configuration
a tight relation between $M_{\rm halo}$ and $M_{200}$ 
is inevitable. The only way $M_{\rm halo} \gg M_{200}$ 
is if the halo has multiple components 
which have been spuriously linked together as
illustrated in the typical example shown in the 
lower panels of Fig.~\ref{fig:fof}.\footnote{These grossly
non-virialized multi-component systems are not always detected 
by more often used relaxation criteria \citep[][and see
Section~\ref{sec:profiles}]{neto07,power12}, as such criteria
focus on the mass within $r_{200}$ which can be in equillibrium
even if diffusely linked to secondary mass concentrations.}
$M_{\rm halo} \ll M_{200}$  could indicate cases where the group
finder has split a virialized object into small fragments. Hence it is
interesting to look at the distribution of $M_{\rm halo}/M_{200}$ for
both the \FoF and \Dhalo algorithms to simply see how $M_{\rm halo}$
compares to the conventional $M_{200}$ definition of halo mass and to
give an indication of the frequency of over linking and
fragmentation.

The two panels of Fig.~\ref{fig:ms2} quantify the distribution of
$M_{\rm halo}/M_{200}$ for both the standard \FOF haloes and for
haloes defined by the \Dhalo algorithm. We immediately see that the
distribution is much tighter for the \Dhalo definition than for \FOF
haloes. For \FOF haloes 5\% of the haloes have $M_{\rm FOF}/M_{200}\gsim 2$ and 1\% $M_{\rm FOF}/M_{200} \gsim 3$.  
In contrast for \Dhaloes only 5\% have $M_{\rm \Dhalo}/M_{200} \gsim 1.5$ and less than
1\%  have $M_{\rm \Dhalo}/M_{200} > 2$. In the \Dhalo panel only \Dhaloes that are
bijectively matched with \FOF haloes are included.  
Since such pairs of haloes contain the same most massive \subhalo, 
the centres used for
calculating $M_{200}$ are identical and result in the same $M_{200}$. 
Furthermore, since
Fig.~\ref{fig:f1} indicates that all \FOF haloes more massive than
$10^{12}$ $h^{-1}$ M$_{\odot}$ have a bijectively matching \Dhalo, then above
$10^{12}$ $h^{-1}$ M$_{\odot}$ we are comparing the same population of haloes
and using the same values of $M_{200}$. Consequently the wider
distribution of $M_{\rm halo}/M_{200}$ for \FOF is directly caused
by the wider spread in $M_{\rm \FoF}$ masses. For the cases where
$M_{\rm \FoF} \gg M_{200}$ there is one or more substantial components
of the \FOF halo that lies outside $r_{200}$. We will see in 
Fig.~\ref{fig:fof} that these are generally secondary mass
concentrations that are linked by tenuous bridges of quite diffuse
material. The \Dhaloes have a tighter distribution of 
$M_{\rm halo}/M_{200}$ as in this algorithm these secondary
concentrations are successfully split off and result in separate
distinct \Dhaloes.

\begin{figure}
\includegraphics[width=8.5cm]{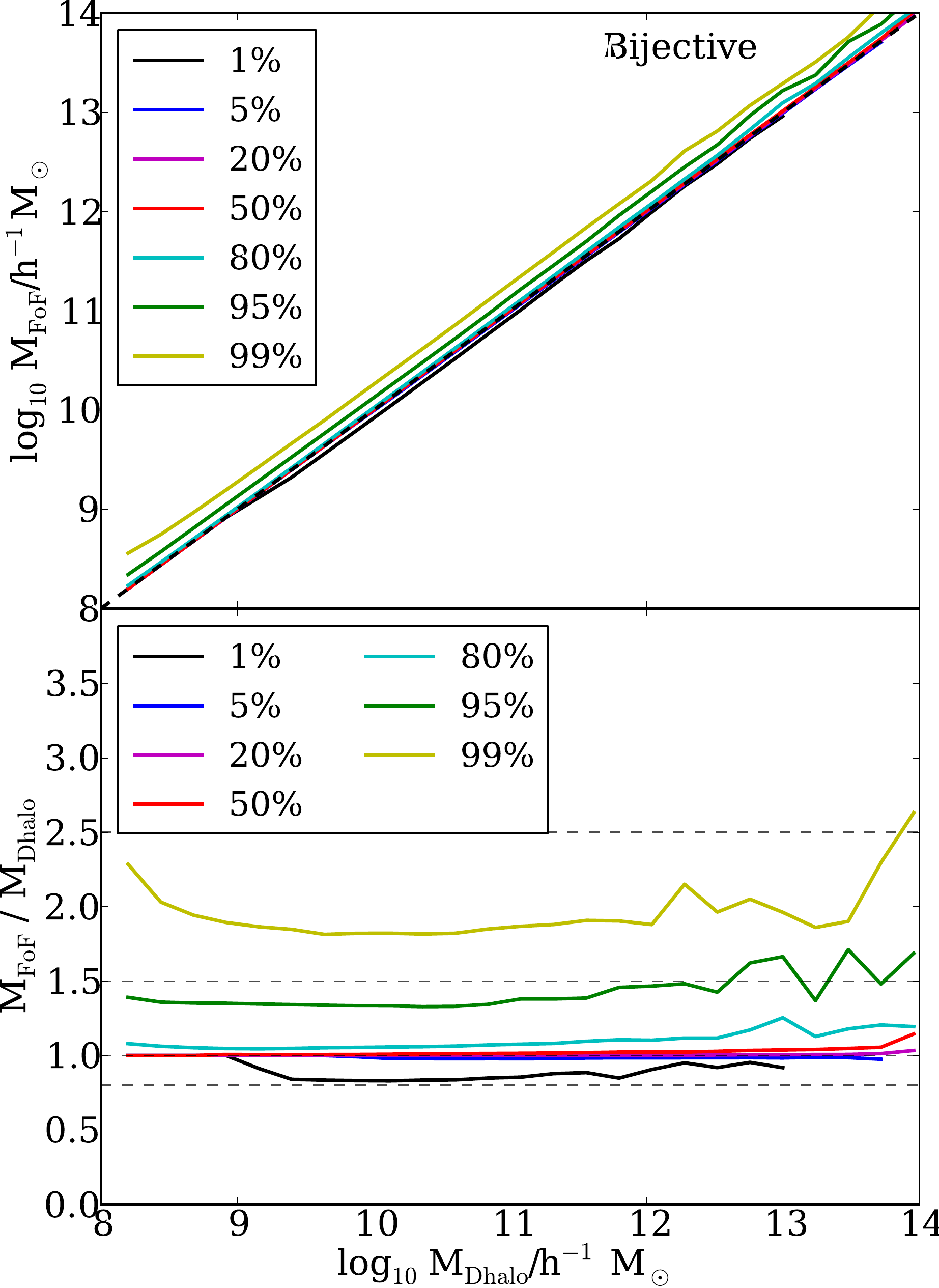}%
\caption{In the top panel, the 1, 5, 20, 50, 80, 95 and 99 percentiles
of the distribution of \FoF halo mass, $M_{\rm \FoF}$, is plotted
against M$_{\rm \Dhalo}$ for the bijectively matched pairs of haloes.
In the bottom panel, the same percentiles of the distribution of the
mass ratio M$_{\rm FoF}$/ M$_{\rm \Dhalo}$ is plotted as a function of
\Dhalo mass. The black dashed lines are where M$_{\rm\FoF}$/M$_{\rm
\Dhalo}$ =0.8, 1, 1.5 and~2.5.}
\label{fig:ms1}
\end{figure}

Our results for \FoF haloes are consistent with earlier
investigations. \cite{harker06, evrard08, lukic09} found that
approximately 80-85\% of \FoF haloes are isolated haloes while
15-20\% of \FoF haloes have irregular morphologies, most of
which are described in \cite{lukic09} as ``bridged haloes". 
The distribution of  $M_{\rm FoF} / M_{\rm 200}$ for ``bridged haloes'' 
given in figure~7 of \cite{lukic09} is very similar to the 
$20\%$ tail of our distribution above 
$M_{\rm FoF} / M_{\rm 200}=1.5$, while the isolated haloes in
\cite{lukic09} have a distribution similar to the remaining 80\%
of our distribution.

\begin{figure*}
\resizebox{6.8cm}{!}{\includegraphics{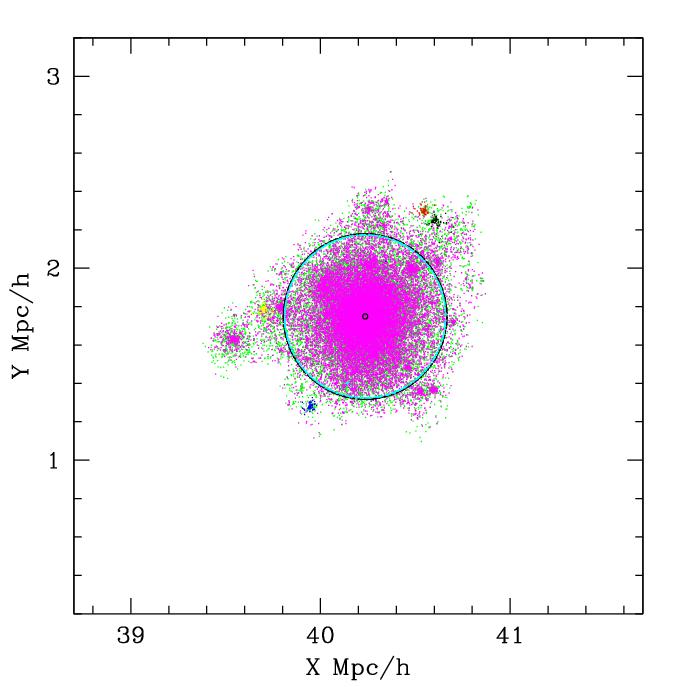}}%
\resizebox{6.8cm}{!}{\includegraphics{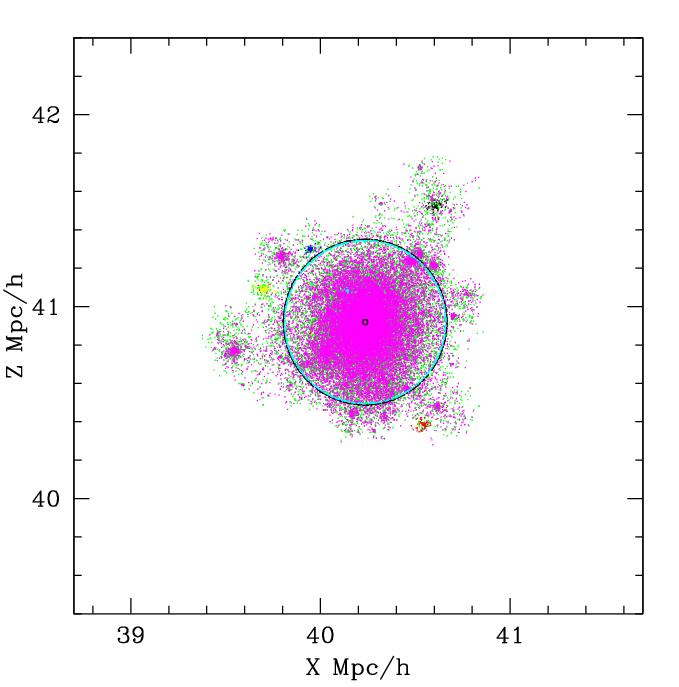}}\\
\resizebox{6.8cm}{!}{\includegraphics{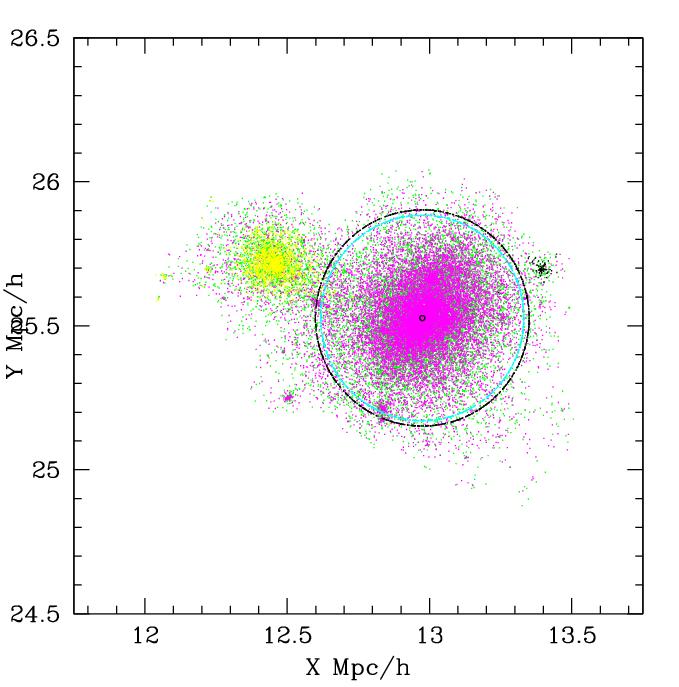}}%
\resizebox{6.8cm}{!}{\includegraphics{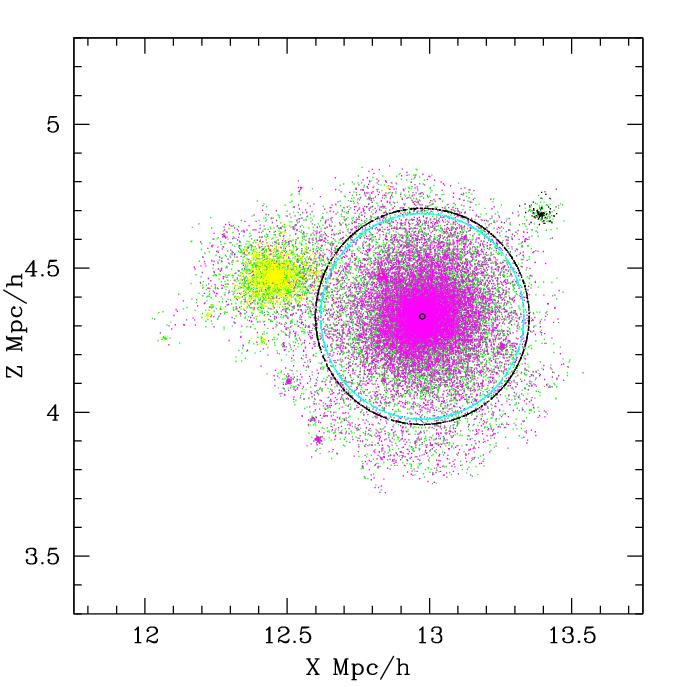}}\\
\resizebox{6.8cm}{!}{\includegraphics{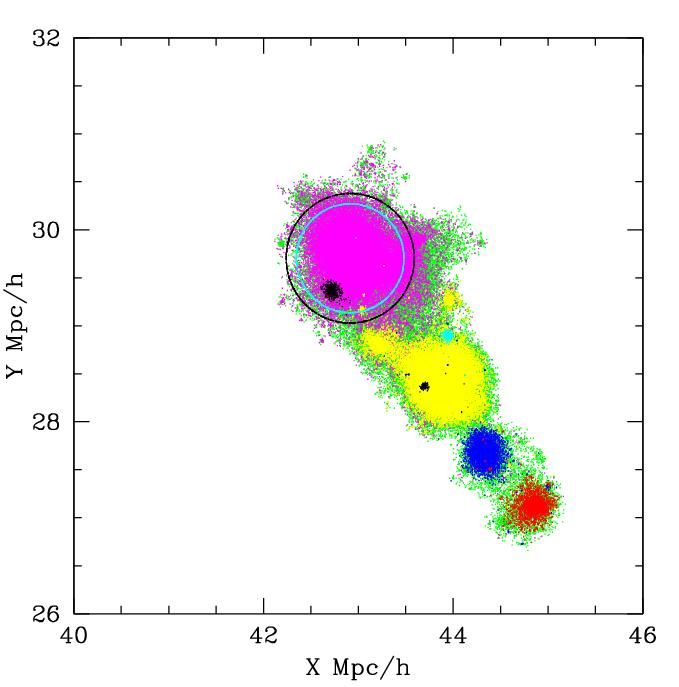}}%
\resizebox{6.8cm}{!}{\includegraphics{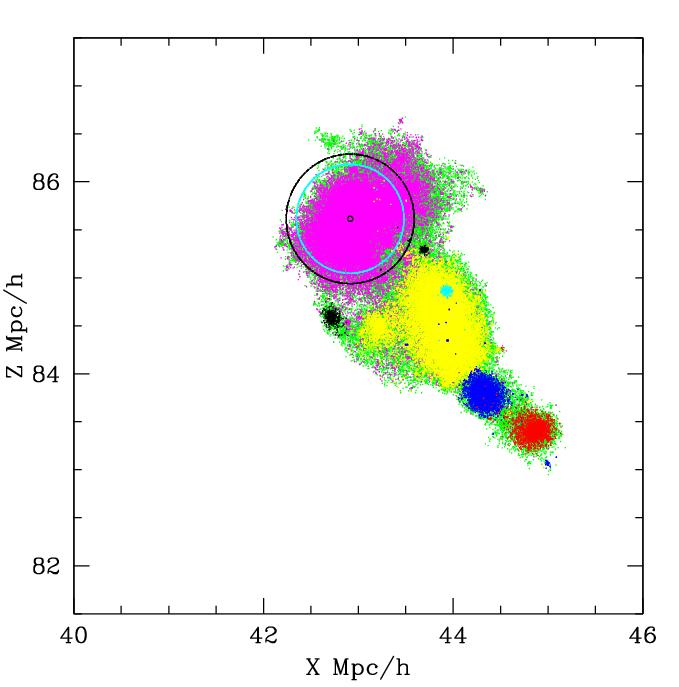}}
\caption{Three examples of the relationship between \FOF haloes and
 \Dhaloes. In each panel all the points plotted are from a single \FoF
 halo. First all the \FOF particles were plotted in green and then
 subsets belonging to specific \Dhaloes were over-plotted.  The magenta
 points are those belonging to the bijectively matched \Dhaloes. Other
 colours are used to indicate particles belonging to other non-bijective 
 \Dhaloes with a unique colour used for each separate \Dhalo. Two projections
 of each halo are shown. The left panels show the X-Y and right the
 X-Z plane. The black circle marks $r_{200}$ of the \FOF halo and
 the cyan circle marks twice the half mass radius of the main \subhalo
 of the \FOF halo.  The top row shows a typical case where $M_{\rm
 FoF} \approx M_{\rm \Dhalo}$.  Here $M_{\rm FoF}= 2.6 \times
 10^{13}h^{-1}$~M$_{\odot}$, $M_{200} = 1.9\times 10^{13} h^{-1}$
 ~M$_{\odot}$, and $r_{200} = 0.43 h^{-1}$~Mpc.  The middle panel
 shows and example where the mass ratio $M_{\rm FoF}/M_{\rm
 \Dhalo}=1.5$ with $M_{\rm FoF} = 1.7 \times 10^{13}
 h^{-1}$~M$_{\odot}$, $M_{200} = 1.2 \times 10^{13}
 h^{-1}$~M$_{\odot}$ and $r_{200} = 0.375 h^{-1}$~Mpc.  The bottom row
 shows an extreme example where $M_{\rm FoF} \gg M_{\rm \Dhalo}$ and
 the \FoF halo is split into many \Dhaloes. Here $M_{\rm FoF} = 1.4
 \times 10^{14} h^{-1}$~M$_{\odot}$, $M_{200} = 7.1\times 10^{13}
 h^{-1}$~M$_{\odot}$ and $r_{200} = 0.67 h^{-1}$~Mpc}
\label{fig:fof}
\end{figure*}
\begin{figure*}
\resizebox{7cm}{!}{\includegraphics{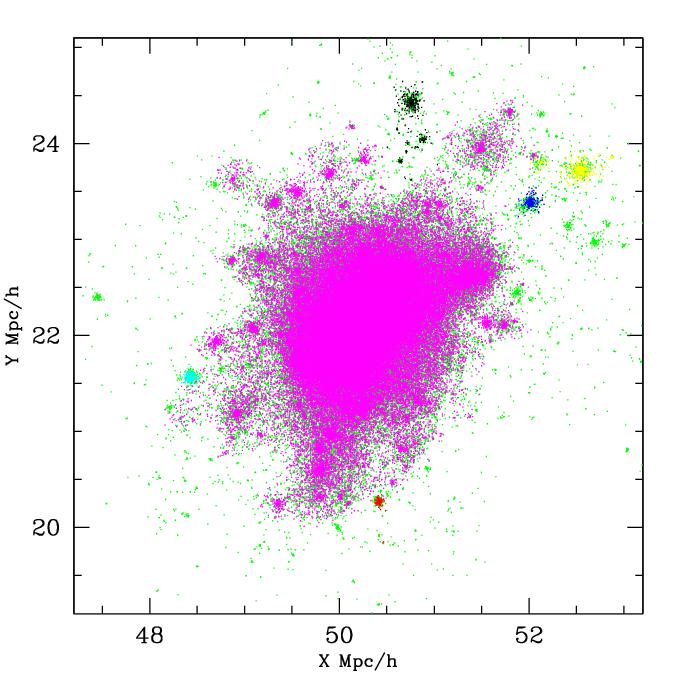}}%
\resizebox{7cm}{!}{\includegraphics{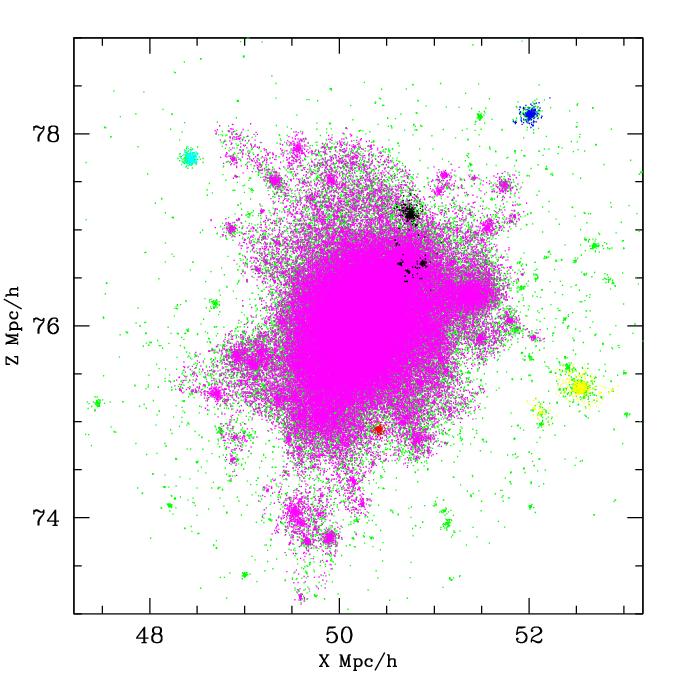}}\\
\resizebox{7cm}{!}{\includegraphics{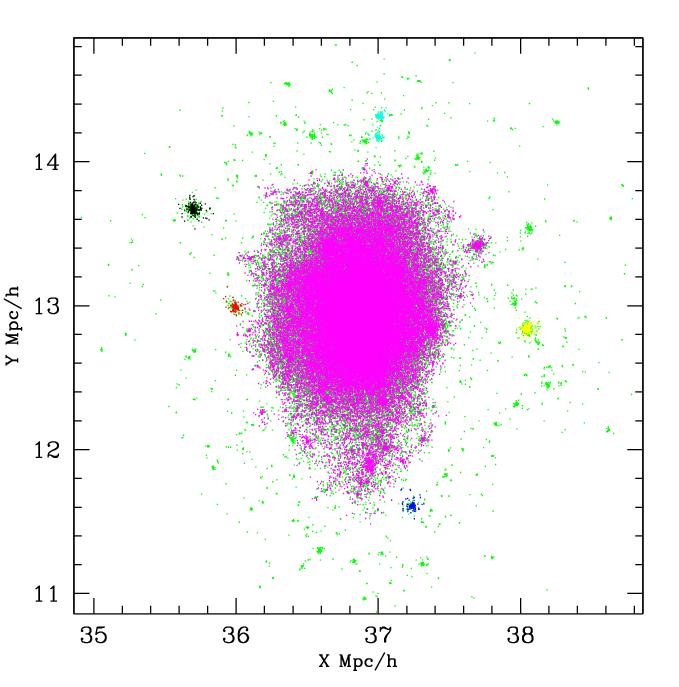}}%
\resizebox{7cm}{!}{\includegraphics{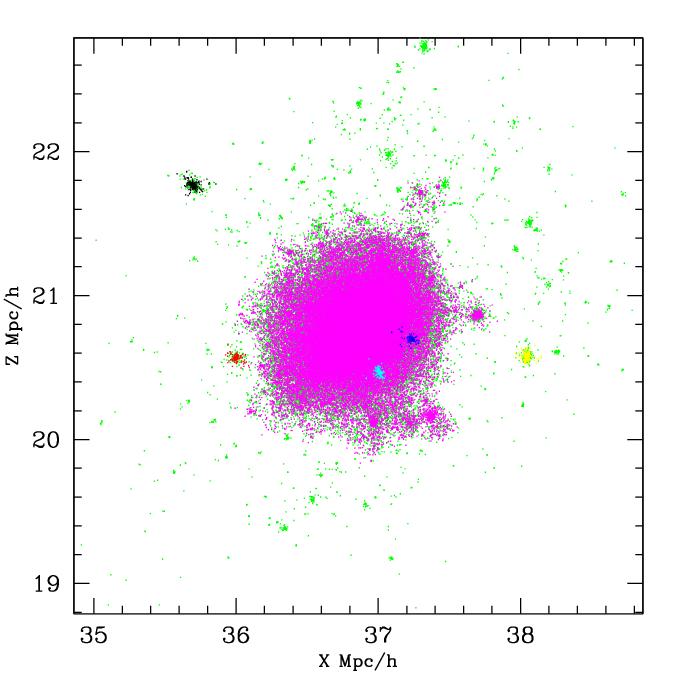}}\\
\resizebox{7cm}{!}{\includegraphics{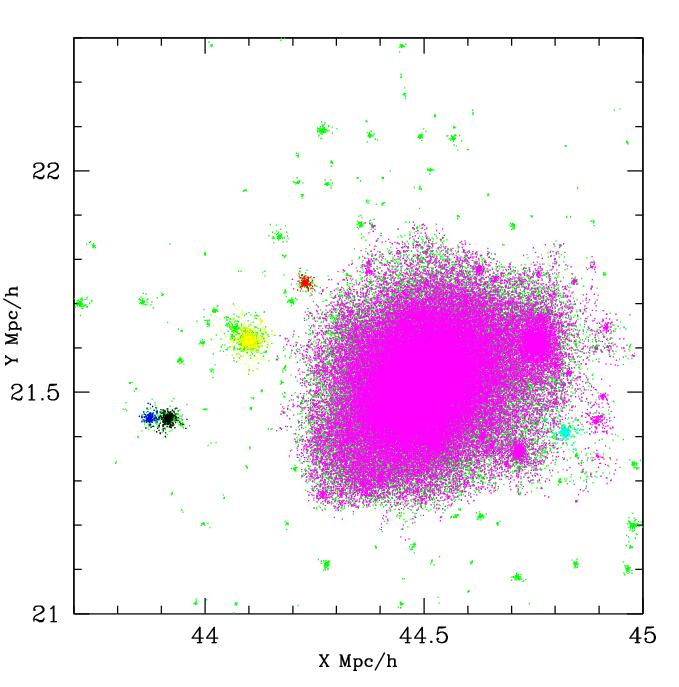}}%
\resizebox{7cm}{!}{\includegraphics{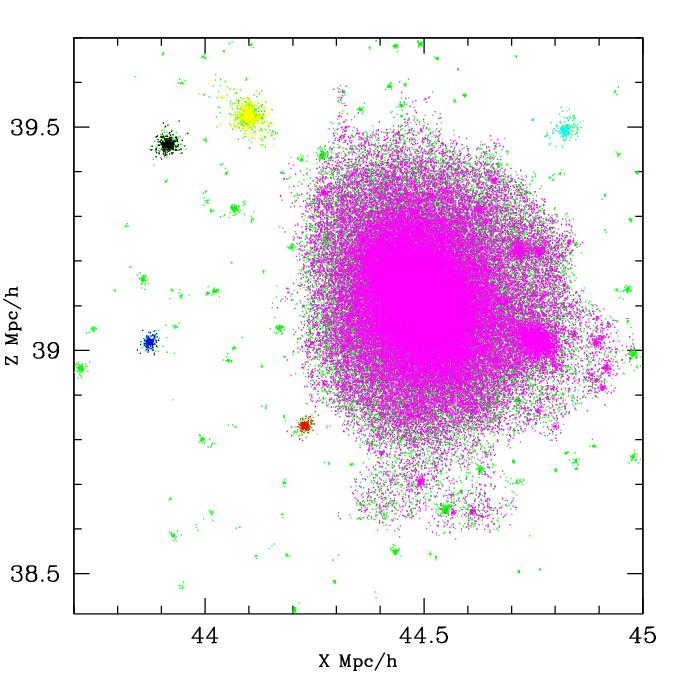}}
\caption{Examples of three typical \Dhaloes showing how a single
\Dhalo can be composed of more than one \FOF halo.  In each panel all
the points plotted are from a single \Dhalo.  First all the \Dhalo
particles were plotted in green and then subsets belonging to specific
\FOF haloes were over plotted.  The magenta points are those belonging to the
bijectively matched \FOF halo. Other colours are used to indicate
particles belonging to other \FOF haloes with a unique colour used for
each separate \FOF halo.  Two projections of each halo are shown. The
left panels show the X-Y and right the X-Z plane.  From top to bottom
the \Dhalo masses of these examples are $M_{\rm \Dhalo} = 4.2 \times
10^{14} h^{-1}$~M$_{\odot}$, $M_{\rm \Dhalo} = 6.8 \times 10^{13}
h^{-1}$~M$_{\odot}$ and $M_{\rm \Dhalo} = 5.4 \times 10^{12}
h^{-1}$~M$_{\odot}$.  In all cases the majority of the \Dhalo mass is
contained in the single bijectively matched \FoF halo and the
secondary \FoF haloes are typically 100 times less massive.  }
\label{fig:dhalo}
\end{figure*}

\subsubsection{Mass Scatter Plots}

We now turn to directly comparing the mass assigned to \FOF haloes and
their corresponding \Dhaloes.  Fig.~\ref{fig:ms1} compares the
distributions of these two masses and their ratio for bijectively
matched \FoF and \Dhaloes, i.e.  haloes which contain the same most
massive \subhalo.  First we see that the median of the distribution is
very close to the one-to-one line. Furthermore on one side the
distribution cuts off very sharply with far fewer than 1\% of haloes
having \FOF masses significantly lower than their corresponding \Dhalo
mass.  In principal $M_{\rm \Dhalo}>M_{\rm FoF}$ can occur as one
aspect of the \Dhalo algorithm is that includes satellite \subhaloes
that previously passed through the main halo even if they are now
sufficiently distant so as not to be linked into the corresponding
\FoF halo. However, such \subhaloes are typically much less massive than
the main \subhalo and the mass gained in this way is out weighed by
other sources of mass loss.  On the other side of the distribution
there is a significant tail of haloes for which $M_{\rm FoF}>M_{\rm
\Dhalo}$. We see that approximately 5\% have $M_{\rm FoF}>1.5 M_{\rm
\Dhalo}$ and 1\% have $M_{\rm FoF}>2 M_{\rm \Dhalo}$.  These fractions
are largely independent of \Dhalo mass.  The main reason for this tail
is the presence of \FoF haloes that have a significant secondary mass
concentration, often linked by a low density bridge, that the \Dhalo
algorithm succeeds in splitting off.  For these bijectively matched
haloes $M_{\rm FoF}$ is unlikely to significantly exceed $2 M_{\rm
\Dhalo}$ as if a single secondary mass concentration had a \subhalo of
mass greater than that of the most massive \subhalo in the \Dhalo we
would not have a bijective match. However, in rare instances $M_{\rm
FoF} > 2 M_{\rm \Dhalo}$ can occur when the \FOF halo contains several
massive secondary mass concentrations.

To illustrate the relationship between \FOF and \Dhaloes we show
three examples in  Fig.~\ref{fig:fof} that have been chosen to
be representative of different points in the 
$M_{\rm FoF}$--$M_{\rm \Dhalo}$ distribution. The halo shown in the
top row is representative of the majority of cases, namely those with
$M_{\rm FoF} \approx M_{\rm \Dhalo}$. Here the only particles from the
\FOF halo that are not included in the \Dhalo are a diffuse cloud of
unbound particles and the particles 
in a couple of \subhaloes whose centres lie outside twice the half mass
radius of the main \subhalo. We stress that these small differences
are what is typical for corresponding \FoF and \Dhaloes.

The middle row of Fig.~\ref{fig:fof} shows an example
where $M_{\rm FoF}/M_{\rm \Dhalo}=1.5$,  which corresponds to the
95th percentile of the distribution shown in Fig.~\ref{fig:ms1}.
Here the \FoF halo is split into three well separated \Dhaloes.
The main \Dhalo is dominant, but there two secondary \Dhaloes, one
a lot more massive than the other, laying outside the $r_{200}$ of the
main \Dhalo. For the purposes of semi-analytic galaxy formation models
such as \GALFORM the three separate haloes given by the \Dhalo
definition is clearly a better description than the single \FoF halo 
as one would not expect the gas reservoirs associated with these
distinct haloes to have merged at this stage and so each should be
able to provide cooling gas to their respective central galaxies.

The bottom row of Fig.~\ref{fig:fof} shows a rare example with $M_{\rm
FoF}/M_{\rm \Dhalo}\approx 2$, the 99th percentile of the
distribution, in which a single \FoF halo is split into several
substantial \Dhaloes. In this and the previous example the \FoF halo
is clearly far from spherical and a large proportion of the \FoF halo
mass lies outside the virial radius that is defined by centring on the
potential minimum of the most massive substructure. Clearly
characterising such haloes by a NFW profile fit just to the mass
within the virial radius would be an inadequate description of
the halo. In fact, in most studies of halo concentrations, including
our analysis present in Section~\ref{sec:profiles}, these haloes would
be deemed to be unrelaxed and excluded from subsequent analysis.  In
contrast, the \Dhaloes in each of the examples presented are much
closer to being spherical with only a small amount of mass outside
their respective virial radii. Each of the primary \Dhaloes in
Fig.~\ref{fig:fof}, including the one in the bottom panel, are
sufficiently symmetrical and virialized to pass the relaxation
criteria that we employ in Section~\ref{sec:profiles} even though the
corresponding \FoF haloes in the bottom two panels are not.

In the example shown in the bottom row of Fig.~\ref{fig:fof} we also
see case of a  \Dhalo that has two distinct components.  Here the two
clumps of black points are a single \Dhalo due to the fact that they
passed directly through each other at a redshift $z=0.89$.  This
extreme example must have been a high speed encounter and so any
galaxies they contained would have been unlikely to merge, but their
extended hot gas distributions would have interacted and possibly
merged. It is for this reason that it is useful in the semi-analytic
models to associate them as a single halo.

\begin{figure*}
\begin{center}
\begin{tabular}{cc}

\resizebox{8cm}{!}{\includegraphics{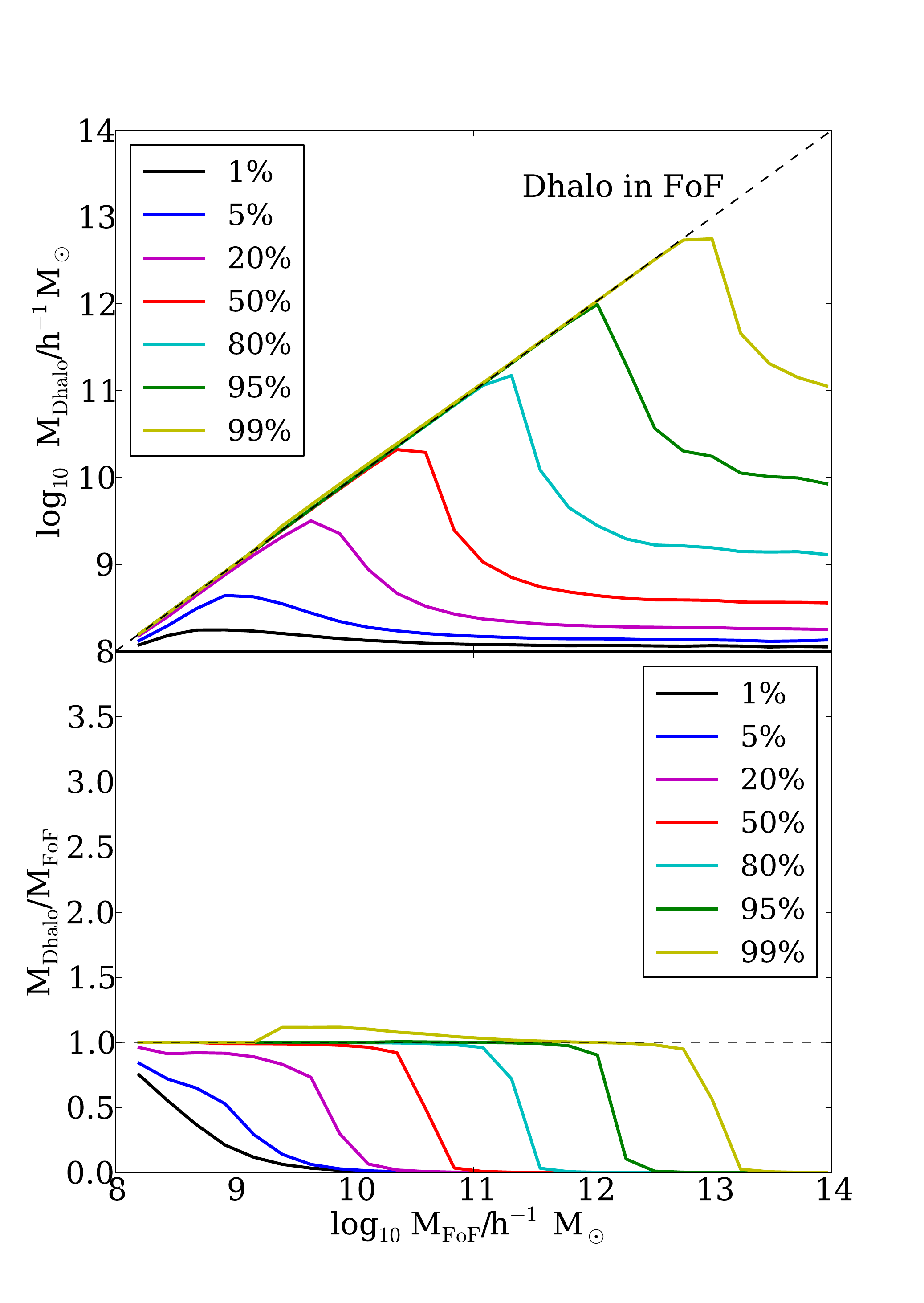}}

\resizebox{8cm}{!}{\includegraphics{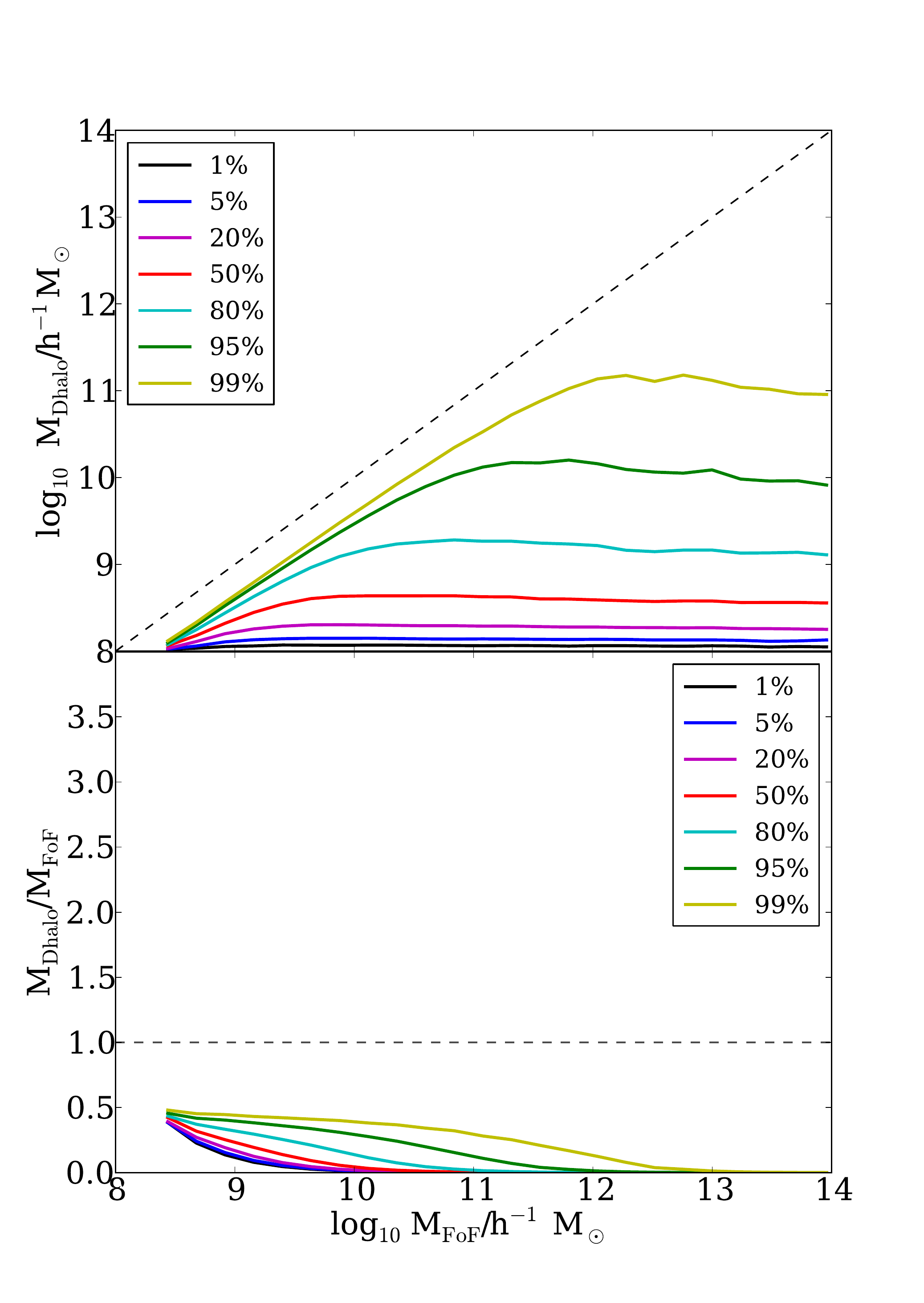}}
\end{tabular}
\end{center}
\caption{In the left hand panels, we plot the median, 1, 5, 20, 80, 95 and
  99 percentiles of the distribution of \Dhalo mass, M$_{\rm Dhalo}$
  (upper), and mass ratio M$_{\rm Dhalo}$/M$_{\rm FoF}$ (lower)
   against M$_{\rm FoF}$ for all the \Dhalo matches to
  each \FoF halo. The black dashed lines in each panel mark
  where M$_{\rm Dhalo}$/M$_{\rm FoF}$ =1. In the right hand panel, we
  plot the same quantities but only for secondary \Dhaloes in each \FoF
  halo.}
 \label{fig:msb}
\end{figure*}

 \begin{figure*}
\begin{center}
\begin{tabular}{cc}

\resizebox{8cm}{!}{\includegraphics{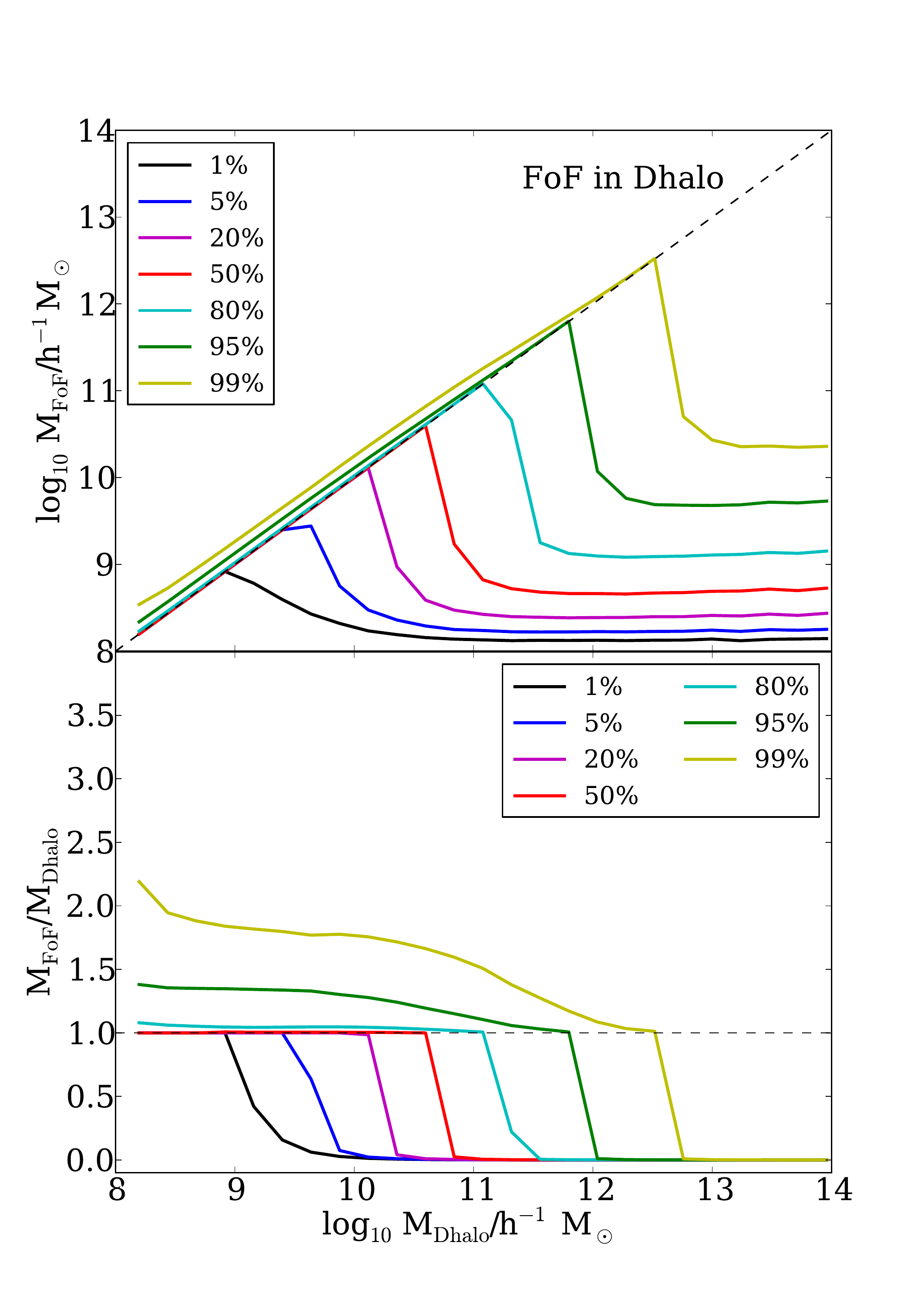}}%
\resizebox{8cm}{!}{\includegraphics{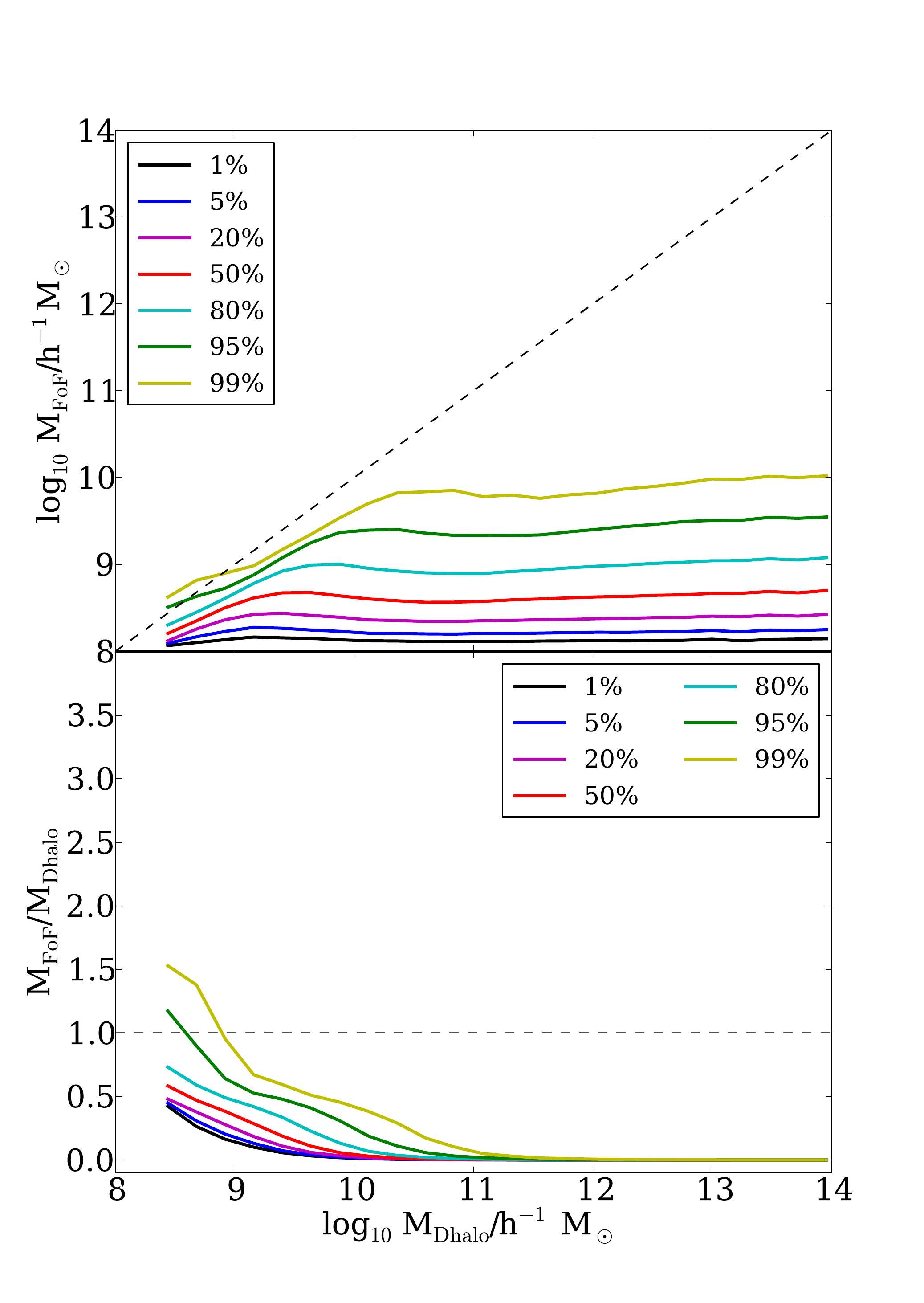}}%
\end{tabular}
\end{center}
\caption{As Fig.~\ref{fig:msb} but with the role of \FOF and \Dhalo reversed.
  In the left hand panels, we plot the median, 1, 5, 20, 80, 95 and
  99 percentiles of the distribution of \FoF halo mass, M$_{\rm FoF}$
  (upper), and mass ratio M$_{\rm FoF}$/M$_{\rm Dhalo}$ (lower)
   against M$_{\rm Dhalo}$ for all the \FoF halo matches to
  each \Dhalo. The black dashed lines in each panel mark
  where M$_{\rm FoF}$/M$_{\rm Dhalo}$ =1. In the right hand panel, we
  plot the same quantities but only for secondary \FoF in each \Dhalo.}
 \label{fig:msa}
\end{figure*}

The \Dhalo algorithm quite frequently merges several \FOF haloes
together into a single \Dhalo as a consequence of the way it avoids
splitting up \subhaloes which at an earlier timestep were in a single
\Dhalo. However unlike the extreme example we have just seen the
typical masses of \subhaloes which pass through a \Dhalo and then
emerge to once again become a distinct \FOF halo are much lower than the
mass of the main \FOF halo. This is illustrated in
Fig.~\ref{fig:dhalo}, where we show the particles of three typical
\Dhaloes of a range of masses colour coded by their \FOF halo membership.
In each case we immediately see that the vast majority of the 
\Dhalo particles also belong to the (bijectively) matched \FOF halo. 
However in addition there are isolated 
clumps of particles in the outskirts of each 
\Dhalo which belong to much smaller distinct \FOF haloes. There are
also similar nearby clumps of particles which due to surrounding
diffuse material are linked into the main \FOF halo. In all cases
each of these clumps are typically less than one percent of the mass
of the main halo.
From the perspective of semi-analytic galaxy formation models
it makes sense to treat each of these clumps equally. For instance, they have 
all been within twice the half mass radius of the main \Dhalo
and could therefore have been ram pressure stripped of their diffuse
gaseous haloes. In \GALFORM satellite galaxies move with the \subhalo
within which they formed 
(or if the descendant of the \subhalo drops below
the 20 particle threshold with the particle that was
previously the potential minimum of its \subhalo)
and so the satellite galaxy positions reflect the spatial distribution
of these \subhaloes even if they move far from the halo to which they are
associated.

\subsection{Non-bijective \FOF and \Dhalo matches}

So far we have just compared \FOF--\Dhalo pairs which form 
a bijective match, that is their most massive \subhaloes are identical.
However there other cases such as the examples of secondary \Dhaloes 
in Fig.~\ref{fig:fof} in which the main \subhalo of the \Dhalo
is not the most massive \subhalo in the corresponding \FOF halo
and conversely examples such as the  secondary \FOF haloes
in Fig.~\ref{fig:dhalo} in which the main \subhalo of the \FOF halo
is not the most massive \subhalo in the corresponding \Dhalo.
We will refer to this former set of matches as \DinF
and the latter as \FinD matches.
Note that the bijective matches are a subset of both of these sets, 
i.e. they are the intersection of the two sets of matches.
To have a complete census of the correspondence between \FOF and
\Dhaloes it is important that we include non-bijectively matched
haloes in our comparison.
We compare the \Dhalo to \FoF halo masses for these two sets
of pairings in Fig.~\ref{fig:msb} and~\ref{fig:msa}. 

The left hand panels of Fig.~\ref{fig:msb} show for all \DinF matches
the dependence of the mass, $M_{\rm \Dhalo}$, and the mass ratio, 
$M_{\rm \Dhalo}/M_{\rm \FOF}$ on the \FOF halo mass. The right hand panel shows
the same quantities but only for secondary \DinF haloes,
i.e. excluding the bijective matches. Focusing first on the right hand
panels, we see that the percentiles of the distribution of secondary
$M_{\rm \Dhalo}$ values are all horizontal lines at high $M_{\rm \FOF}$,
indicating that in this regime the distribution of $M_{\rm \Dhalo}$ is
independent of $M_{\rm \FOF}$. This suggests that the secondary \Dhaloes
that are linked into high mass \FOF haloes by bridges of diffuse
material are essentially drawn at random from the \Dhalo
population. We note that in this way the \FOF halo can be hundreds or
more times more massive than many of the \Dhaloes in contain.  In
these same panels, we see that at lower masses the distribution of
\Dhalo masses is sharply truncated at $M_{\rm \Dhalo} = M_{\rm \FOF}/2$. This
is essentially by construction as if a \Dhalo with mass greater than
$M_{\rm \FOF}/2$ were linked into the \FOF halo then its most massive
\subhalo would very likely to be the most massive \subhalo of the whole
\FOF halo and hence there would be a bijective match and this pairing
would be excluded from this plot.  The left hand panels of
Fig.~\ref{fig:msb}, which includes the bijective matches show a more
complex distribution. However it can be easily understood as resulting
from the superposition of the distribution from the right hand panel
with the distribution of bijective matches shown in Fig.~\ref{fig:ms1}. At
very low masses most \FOF haloes contain only a single resolved
\subhalo and so the \FOF halo cannot be split into multiple \Dhaloes
and so the overall distribution is dominated by the bijective matches
resulting in a tight correlation between $M_{\rm \Dhalo}$ and $M_{\rm \FOF}$.
With increasing \FOF mass there are more and more secondary \Dhaloes
per \FOF halo. They increasingly dominate over the bijective matches
and so the contours tend to their values in the right hand panel.

\begin{figure}
\begin{center}
\begin{tabular}{cc}
\resizebox{8cm}{!}{\includegraphics{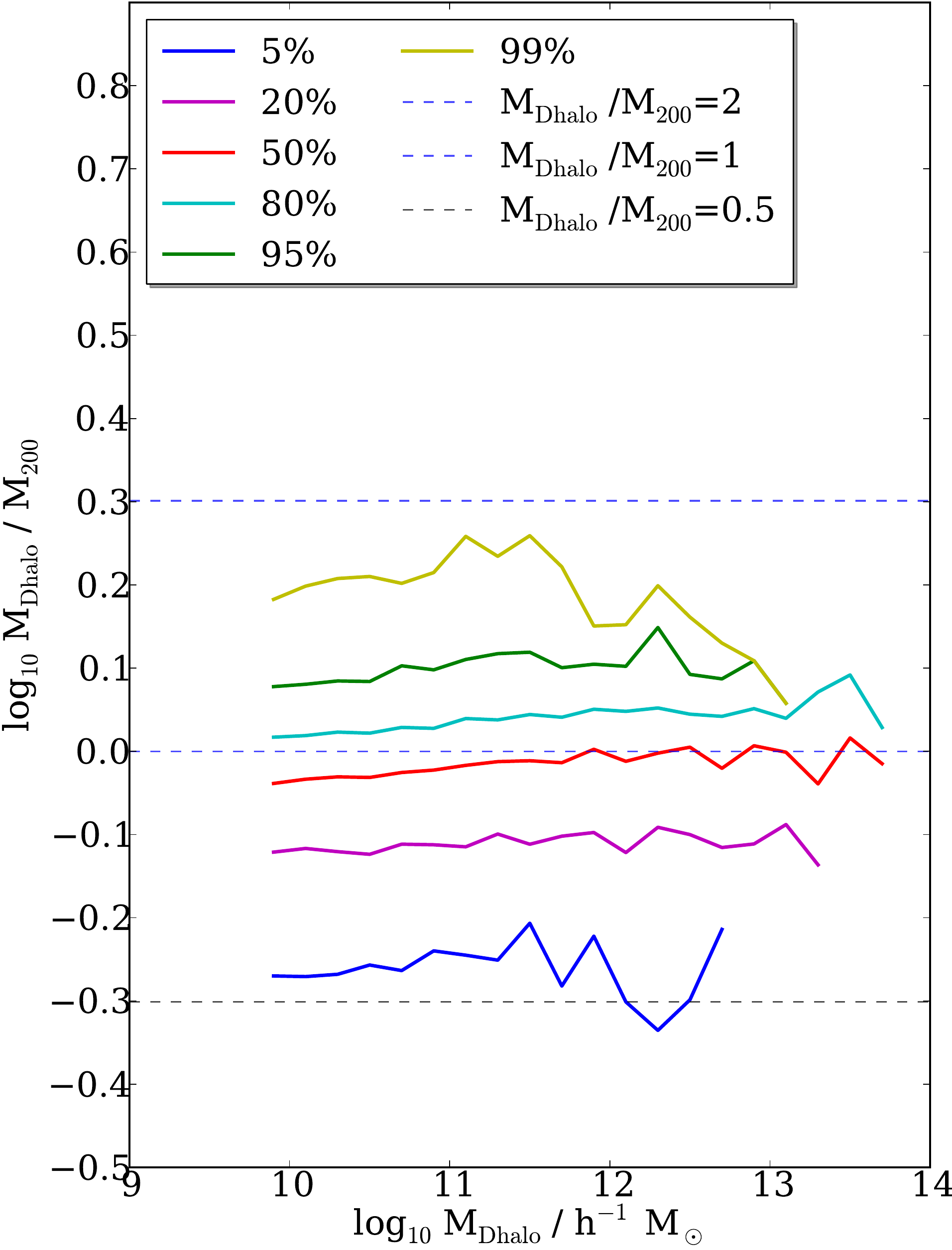}}%
\end{tabular}
\end{center}
\caption{ Like the right hand panel of Fig.~\ref{fig:ms2}, but for
non-bijective \Dhaloes. The curves show the median, 5, 20, 80, 95, 99
percentiles of the ratio between the \Dhalo mass, M$_{\rm Dhalo}$, and the
virial mass, M$_{200}$.  The horizontal dashed lines indicate M$_{\rm
Dhalo}$/M$_{\rm 200}$ = 0.5, 1.0, 2.0.}
\label{fig:nonbi_distrb} 
\end{figure}
\begin{figure}
\resizebox{8.5cm}{!}{\includegraphics{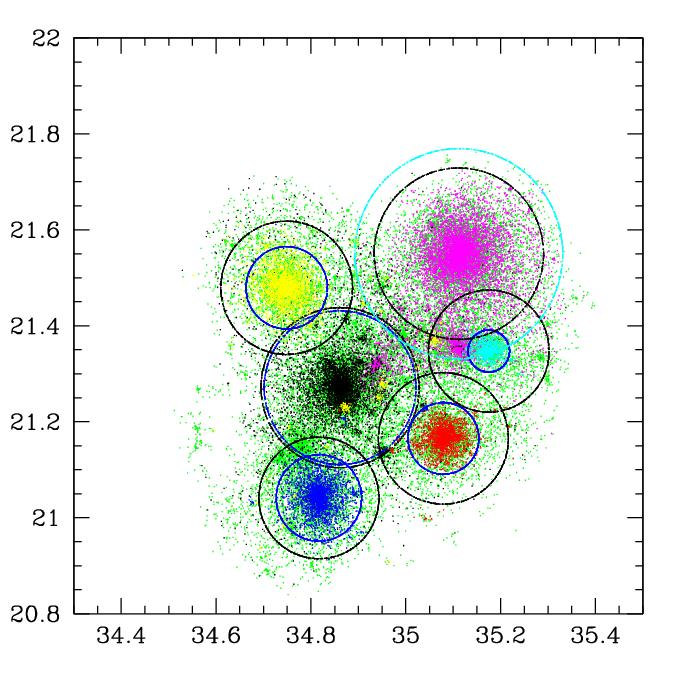}}%
\caption{An example of one \FOF halo split by the \Dhalo algorithm
into several \Dhaloes.  All the points plotted are from a single \FoF
halo. First all the \FOF particles are plotted in green and then
subsets belonging to specific \Dhaloes are over plotted.  The magenta
points are those belonging to the bijectively matched \Dhalo. Other
colours are used to indicate particles belonging to other \Dhaloes
with a unique colour used for each separate \Dhalo.  The black circle
around the magenta points marks  $r_{200}$ of the \FoF halo and is
also the $r_{200}$ of the bijective \Dhalo. The concentric cyan circle
marks twice the half mass radius of this main \subhalo.
The other black circles show $r_{200}$ locations for the non-bijective
\Dhaloes, while the concentric blue circles indicate twice the 
half mass radius of the corresponding \subhalo. }
\label{fig:nonbi}
\end{figure}

Fig.~\ref{fig:msa} shows the distribution of \FOF halo mass for the
\FinD matches. Again the right hand panes show the distribution for
just the secondary matches while the left hand panels also include the
primary or bijective matches. Comparing the right hand panels of
Fig.~\ref{fig:msa} and Fig.~\ref{fig:msb} we see that the
corresponding contours are shifted to lower masses. Thus it is rarer
for a \Dhalo to contain massive secondary \FOF halo than it is for
\FOF halo to contain massive secondary \Dhalo. The secondary \Dhaloes
arise from the remerging step in the \Dhalo algorithm whereby two
\subhaloes that have passed through each other (the smaller has come
within twice the half mass radius of the larger) are deemed thereafter
always to be part (or satellite components) of the same \Dhalo even if
they subsequently separate sufficiently to become distinct \FOF
haloes.  This occurs reasonably frequently, but as in the examples
shown in Fig~\ref{fig:fof} the secondary \FOF haloes are typically
much less massive than the primary and contribute little to the total
mass of the halo.  Interestingly the near horizontal contours in the
upper right hand panel Fig.~\ref{fig:msa} indicate that the mass
distribution of this population of secondary \FOF haloes is
approximately independent of $M_{\rm \Dhalo}$ for high \Dhalo
masses. As these \FOF haloes are often heavily stripped by their
passage through the main \Dhalo this is not a trivial result. The
contours begin to dip at lower masses reflecting the fact it is
unlikely for a matched \FOF halo to have a mass greater than about one
half of $M_{\rm \Dhalo}$ without it being the primary or bijective
match.  This expectation is violated for $M_{\rm
\Dhalo}<10^9h^{-1}$~M$_\odot$, but this is a resolution effect because
at such low masses secondaries with $M_{\rm FOF}\ll M_{\rm \Dhalo}$
fall below the 20 particle limit of the catalogue and so their absence
biases the distribution towards higher ratios.

The left hand panels of Fig.~\ref{fig:msa} are for all the matches of
\FinD, including the bijective matches. These distributions can be
understood as a superposition of the distributions in the right hand
panels with the distribution for bijective matches shown in
Fig.~\ref{fig:ms1}. At low masses the bijective halo matches dominate
whereas at large $M_{\rm \Dhalo}$ there are many \FOF haloes matched
to each \Dhalo.  Thus, for example, at $M_{\rm \Dhalo}\approx
10^{10.5}h^{-1}$~M$_\odot$ we transition from 50\% of the matched \FOF
haloes being primary to 50\% of them being much lower mass ($M_{\rm
\FoF}\approx 10^{8.7 }h^{-1}$~M$_\odot$) secondary \FOF haloes.

In section 3.1.1, we examined the distribution of the $M_{\rm \Dhalo
}/M_{\rm 200}$ ratio for the bijectively matched haloes. We are also
interested this distribution for the non-bijective \Dhaloes shown in
Fig.~\ref{fig:nonbi_distrb}.  We immediately notice the distribution
is shifted towards lower values than the corresponding distribution
for the bijective haloes shown in Fig.~\ref{fig:ms2}.  The origin of this
shift can be understood by reference to Fig.~\ref{fig:nonbi} which
shows an example of a \FoF halo which is split into several \Dhaloes.
The \Dhalo whose particles are plotted in magenta is the bijective
match of the \FoF halo and the \Dhaloes plotted in other colours are
non-bijective matches.  The black circles in Fig.~\ref{fig:nonbi}
show the location of $r_{200}$ for each of the \Dhaloes, while the
other circles show the location of the half-mass radius of each
\Dhalo. For bijectively matched \Dhaloes, the majority of which are
isolated, $r_{200}$ is typically slightly smaller than the half-mass
radius.  In contrast we see in Fig.~\ref{fig:nonbi} that for many of
the non-bijectively matched \Dhaloes the half mass radius is much
smaller than $r_{200}$.  This is a consequence of the \subfind
algorithm which determines the extent of a \subhalo by finding saddle
points in the density distribution \citep{springel01}. Hence as a
\subhalo enters a dense environment the mass assigned to it by 
\subfind is decreased. This environmentally dependent effect both
lowers $M_{\rm \Dhalo}$ relative to $M_{\rm 200}$ and increases the
scatter in this relation.

\section{Statistical Properties of \Dhaloes}
\label{sec:stat}

Having thoroughly compared individual \Dhaloes with their
corresponding \FOF haloes,
we now turn to the statistical properties of the \Dhaloes.
We first look at the \Dhalo mass function and then the statistics
of their density profiles as characterised by fitting NFW 
profiles \citep{nfw95,nfw96,nfw97}.

\subsection{The \Dhalo mass function}

For many applications it is extremely useful to have an analytic
description of the number density of haloes as a function of halo mass.
A relevant example for us is when semi-analytic galaxy formation
models are constructed using Monte-Carlo methods  \citep{parkinson08, cole00}
of generating dark matter merger trees. In this case, in order to
construct predictions of galaxy luminosity functions or any other
volume averaged quantity\citep{cole00,berlind03,baugh05,neistein08,bundy05,giocoli08,moreno08,van05}, one needs knowledge 
of the halo mass function in order to know how many of each type of
tree one has per unit volume. It has become common practice
to assume the halo mass function is given by analytic
fitting functions 
which have been fitted to the abundance of haloes found by the \FoF or
other group finding algorithms \citep{davis85, lacey94,kk09} in suites
of cosmological N-body simulations.  \cite{murray13a} compare all the
currently proposed fitting functions.
 In our semi-analytic
modelling we would like to achieve consistent results when using 
Monte-Carlo merger trees or when using merger trees extracted
directly from N-body simulations using the \Dhalo algorithm. Hence it
is important to directly determine the \Dhalo mass function and to
compare it to such fitting formulae.

\begin{figure*}
\includegraphics[width=8cm]{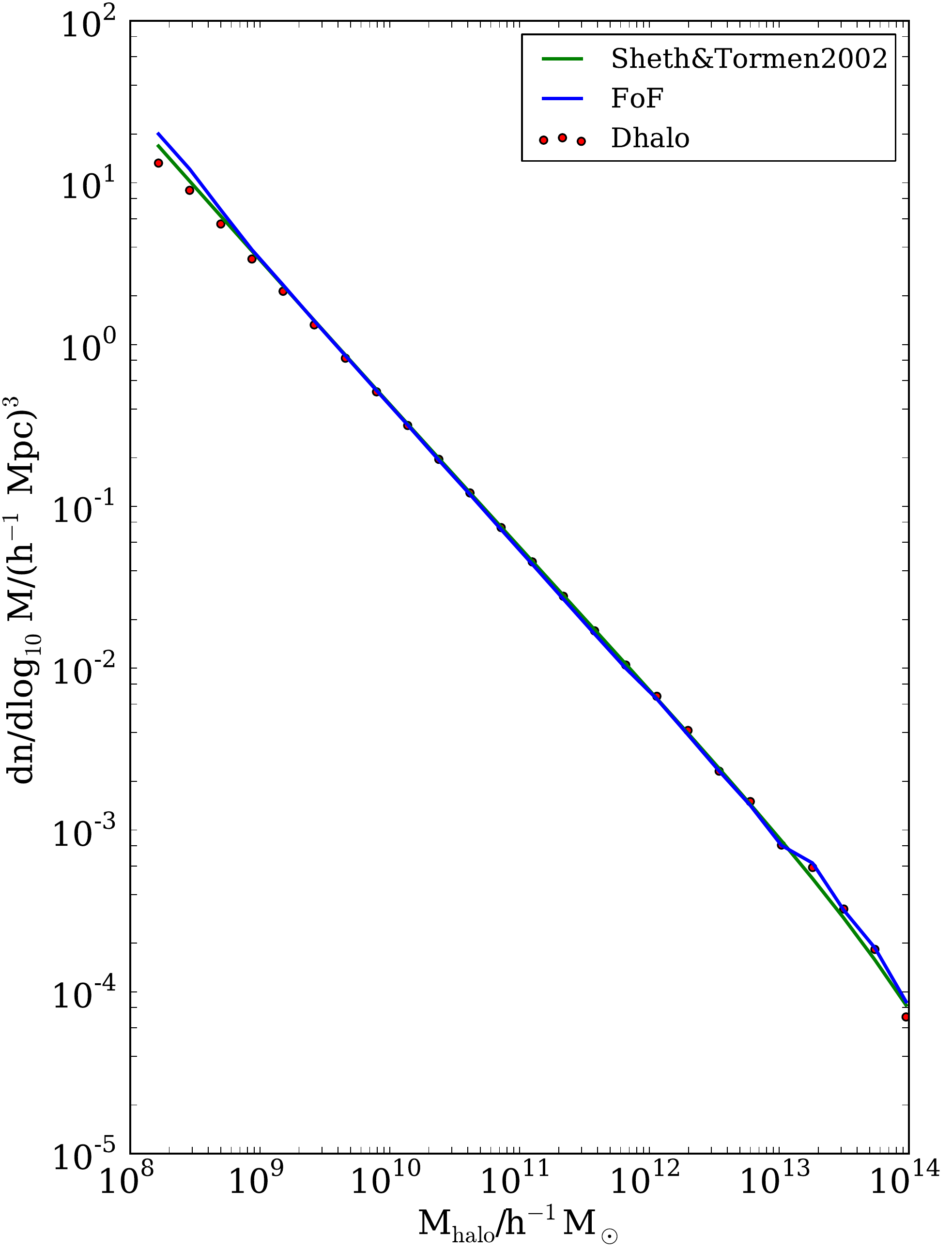}%
\includegraphics[width=7.7cm]{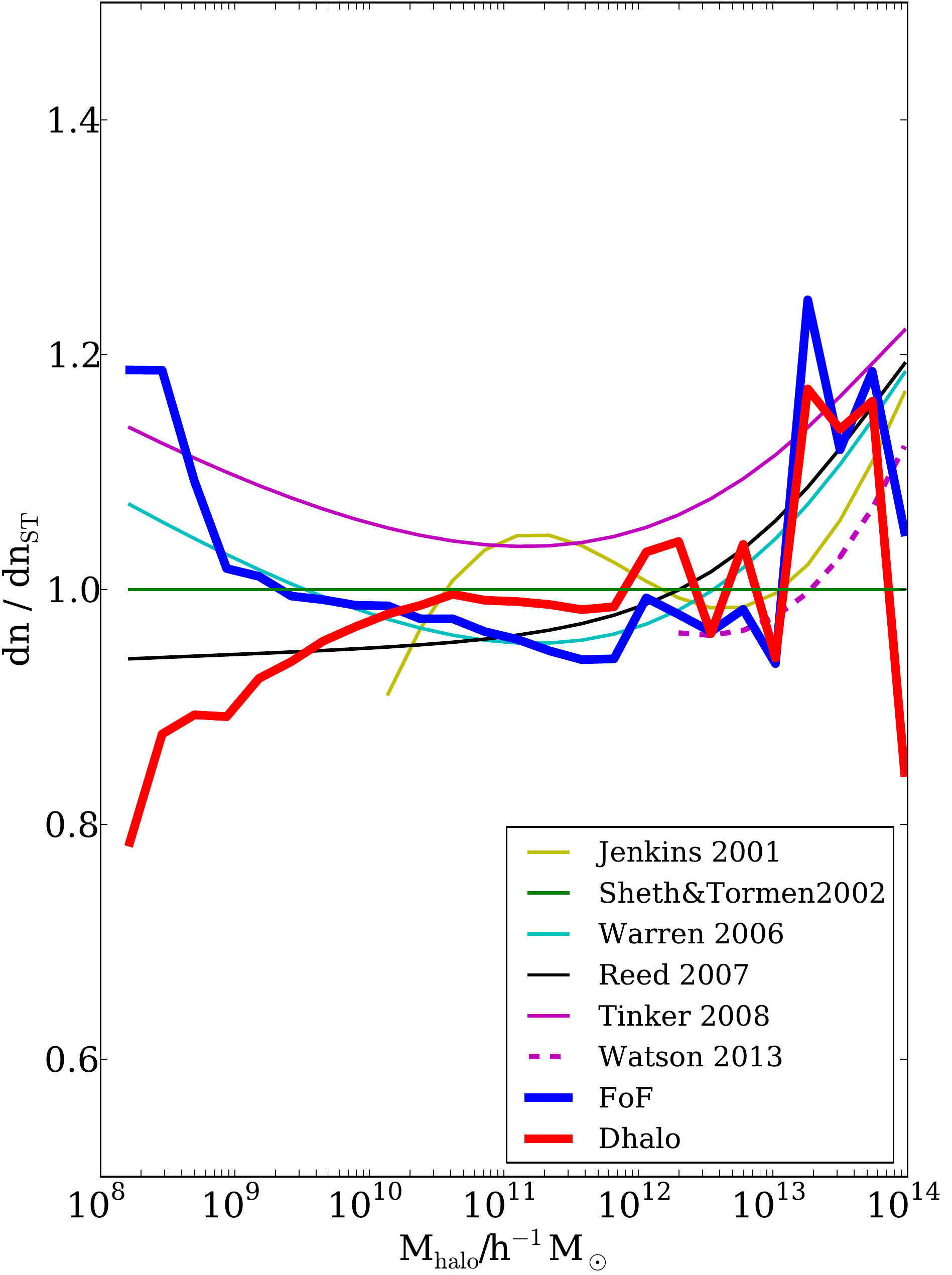}%
\caption{The left hand panel shows the differential mass functions for
both \FOF (linking length $b=0.2$) haloes (blue line) and  \Dhaloes
(red points) in the MSII simulation. We plot this down to $\sim
10^8h^{-1}$~M$_\odot$, the mass corresponding to 20 particles in the MSII
simulation and we also plot the Sheth and Tormen (2002) mass function as a comparison. To expand the dynamic range, the right hand
panel shows  the corresponding prediction of various analytic mass functions(Jenkins et al 2001,Warren et al 2006, Reed et al 2007, Tinker et al 2008, Watson et al 2013) as indicated in the legend but now relative to the Sheth and Tormen(2002) prediction. The \FOF 
\Dhalo data are now shown as the heavy blue and red
lines. }
\label{fig:halomass}
\end{figure*}

We do this in Fig.~\ref{fig:halomass} which compares the \Dhalo and
\FoF mass functions that we measure in the MSII simulations with
various analytic prescriptions
\citep{jenkins01, sheth02, warren06, reed07,tinker08,watson13}.
The left hand panel shows the number
density of haloes per unit logarithmic interval of mass from the
nominal 20 particle mass resolution of the simulation up to
$10^{14}h^{-1}$~M$_\odot$ which is the mass of the most massive haloes
in the simulation.  In constructing these mass functions the
halo mass we use is simply the aggregated mass of all the particles
assigned to each halo. Thus in the \FOF case this is all particles
linked to the halo by the \FOF algorithm while in the \Dhalo case it
is the
sum of the masses of the \subhaloes that compose an individual \Dhalo.
Also shown on this panel are the predictions of various analytic
prescriptions. To evaluate these we use $\sigma^2(M)$, the variance of
the density fluctuations as a function of mass (using a top hat
filter), corresponding to the input power spectrum of the MSII
propagated to the output time of the simulation using linear theory.
They are all clearly very similar and so in the left hand panel we
expand the dynamic range of the comparison by plotting each mass 
function divided by the prediction of the Sheth \& Tormen (2002)
model.

The first thing that we note is that despite the sometimes quite large
differences (see \S\ref{sec:single}) in the masses of individual \FOF
and \Dhaloes their two mass functions agree to within 5\% for all
masses greater than $10^{10}h^{-1}$~M$_\odot$. In the range $10^{10}
\lsim M_{\rm halo} \lsim 10^{12.5} h^{-1}$~M$_\odot$ the \Dhalo
abundance is approximately 5\% higher than \FOF haloes as roughly 5\%
of \Dhaloes are secondary members of \FOF haloes.  
 In other words, the \FOF halo abundance has been suppressed relative
 to the \Dhalo abundance by a fraction of them being composed of two
 or more \Dhaloes that have been
linked into one more massive \FOF halo by diffuse material or bridges.
There is also a competing effect, \FOF haloes being remerged into
 single \Dhaloes, which suppressed the \Dhalo abundance, but this is a
 much smaller effect.

Below $10^{10}h^{-1} $~M$_\odot$ the abundance of \FOF halo rises
systematically above that of \Dhaloes. 
Between $10^{10}h^{-1}$~M$_\odot$ 
and $8 \times 10^{8}h^{-1}$~M$_\odot$ this excess increases to about
10\% and is caused by 
\FOF haloes that are remerged to become secondary  components of a
larger \Dhaloes (see Fig.~\ref{fig:f1}).  
At lower masses (\lsim 100 particles) the sharp up turn  in the \FOF mass function
relative to that of \Dhaloes is due to an increasing 
 fraction of the \FOF haloes not containing a self-bound
\subhalo and so having no corresponding \Dhalo  (see Fig.~\ref{fig:f1}).
Thus this portion of the
mass function is strongly affected by the  resolution of the simulation.

The Jenkins et al (2001) fitting formula is within 10\% of both the
\FOF and \Dhalo mass functions for masses above $2\times
10^{10}h^{-1}$~M$_\odot$. However below this mass it strongly
under predicts the number density of low mass haloes.  Note that we
only plot this fit and that of Watson et al (2013)
over the mass ranges used to constrain them in the original papers. 
The Watson et al (2013) mass function is only defined at very high
masses where we have poor statistics. It lies somewhat below but is
still compatible with our noisy estimates.
The Warren (2006) model has the best agreement
with our \FOF mass function, fitting it well all the way down to 40
particles, beyond which we expect our limited resolution means that our
\FOF mass function is contaminated by spurious unbound chance
groupings of particles. 
However the Reed (2007) mass function does a
better job of matching the low mass end of our \Dhalo mass
function. The Sheth \& Tormen mass function is intermediate at low
masses between that of Warren (2006) and Reed (2007), but
systematically below the other models and our \FOF and \Dhalo mass
function at high masses, though still only at the 15\% level.
The Tinker (2008) mass function predicts halo abundances that are 
about 5 to 10\% higher than Warren (2006) and our estimated \FOF abundances.

In summary, the \Dhalo and \FOF mass functions are very similar and
only differ by more than 5\% below $10^{10}h^{-1}$~M$_\odot$.  As a
result the established analytic mass function models fit the \Dhalo
mass function almost as well as they do the standard \FOF mass
function. The differences between the different analytic fitting
formulae are greater than the difference between the \FOF and \Dhalo mass 
functions. 
The Reed (2007) model is a slightly better
description of the \Dhalo mass function due to it predicting a
slightly lower abundance at low masses.

\subsection{Density Profile Fits}
\label{sec:profiles}

We now turn to the density profiles of the haloes as these are an
important ingredient in semi-analytic models such as \GALFORM where they
influence the rate at which gas cools and set the gravitational
potential well in which galaxies form.  We choose to fit the halo
density profiles using NFW  \citep{nfw96,nfw97} profiles
\begin{equation}
\frac{\rho_{\rm NFW}(r)}{\rho_{\rm crit}}  
= \frac{\delta_c}{{r}/{r_{\rm s}} (1+
  {r}/{r_{\rm s}})^{2} }\quad (r\leq r_{200}), 
\end{equation}
where $\delta_c$ is the characteristic density contrast, and $r_{\rm s}$
is the scale radius. We define the virial radius, $r_{200}$, as the radius at
which the mean interior density equals $200$ times the
critical density, $\rho_{\rm crit} = 3H_{0}^{2}/(8\pi G)$. The
concentration is defined as $c \equiv r_{200}/r_{\rm s}$. The definition of
$r_{200}$ implies that $\delta_c$ and $c$ must
satisfy 
\begin{equation}
\delta_c  = \frac{200}{3}\frac{c^{3}}{\ln(1+c)-{c}/{(c+1)}} .
\end{equation}

Our choice of NFW profiles is motivated by their accuracy as a model
of CDM haloes \citep{nfw96,nfw97}, their
widespread use and so that our results can be compared to those in
\cite{neto07}
who studied the statistics of NFW concentrations for \FOF haloes
identified in the Millennium Simulation \citep{springel05a}. To allow us to
compare directly with \cite{neto07} we have followed their fitting procedure.

For each halo, we have computed a spherically-averaged density profile
by binning the halo mass into 32 equally spaced bins in $\log_{10}(r)$
between the virial radius and $\log_{10}(r/r_{200}) = -2.5$, centred
on the potential minimum.
We fit the two free parameters, $\delta_c$ and $r_{\rm s}$ by minimising
the mean square deviation
\begin{equation}
\sigma_{\rm fit}^{2}  = \frac{1}{N_{\rm bin}-1} \sum_{i}^{N_{\rm
    bin}}[\log_{10} \rho(r_{i}) - \log_{10}\rho_{\rm NFW}(r_i\vert
    \delta_c,r_{\rm s})]^2
\end{equation}
between the binned $\rho(r)$ and the NFW profile. 
As in \cite{neto07}, we perform the fit over the radial range $0.05<r/r_{200}<1$.
In order to be consistent with the original NFW work, we express the
results in terms of fitted virial mass, $M_{200}$, and a concentration, 
$c_{200} \equiv r_{200}/r_{\rm s}$.  We note that while the fitted
value of $M_{200}$ used here and the directly measured $M_{200}$ used
earlier (e.g. in Fig.~\ref{fig:ms2}) are not identical they in general
agree very accurately with an rms scatter of less than 3\%.

\cite{neto07} distinguished relaxed haloes from haloes that were not
in dynamical equilibrium due to recent or ongoing mergers. They found
that relaxed haloes were well fit by NFW profiles while the profiles
of unrelaxed haloes were lumpier and yielded poorer fits with
systematically lower concentrations. Hence to compare to \cite{neto07}
we use the following three objective criteria to assess 
whether a halo has reached
equilibrium \cite{neto07,gao08,power12}.
 \begin{enumerate}
\item
 The fraction of mass in resolved substructures whose centres
 lie inside $r_{200}$: $f_{\rm sub} =\sum_{i\neq 0}^{N_{\rm sub}}
 M_{{\rm sub},i} / M_{200}$. We require $f_{\rm sub}<0.1$ for relaxed
 haloes. 
\item The
 centre of mass displacement, i.e. the difference between the position
 of the potential minimum  and the centre of mass, $s =
 |r_{\rm c}-r_{\rm cm}|/r_{200}$~\citep[]{thomas01}. Note that,
 the centre of mass is calculated using  all the particles within $r_{200}$, not only those
 belonging to the \FoF  or \Dhalo. We require $s <  0.07$ for relaxed haloes.
\item The virial ratio, $2T/|U|$, where $T$ is
 the total kinetic energy of halo particles within $r_{200}$ and $U$
 is their gravitational potential self energy. We require
 $2T/|U|<1.35$ for our relaxed haloes.  (For haloes with more than 5000
 particles we use a random subset of 5000 particles to estimate $U$.)
 \end{enumerate}

\begin{figure}
\includegraphics[width=8.5cm]{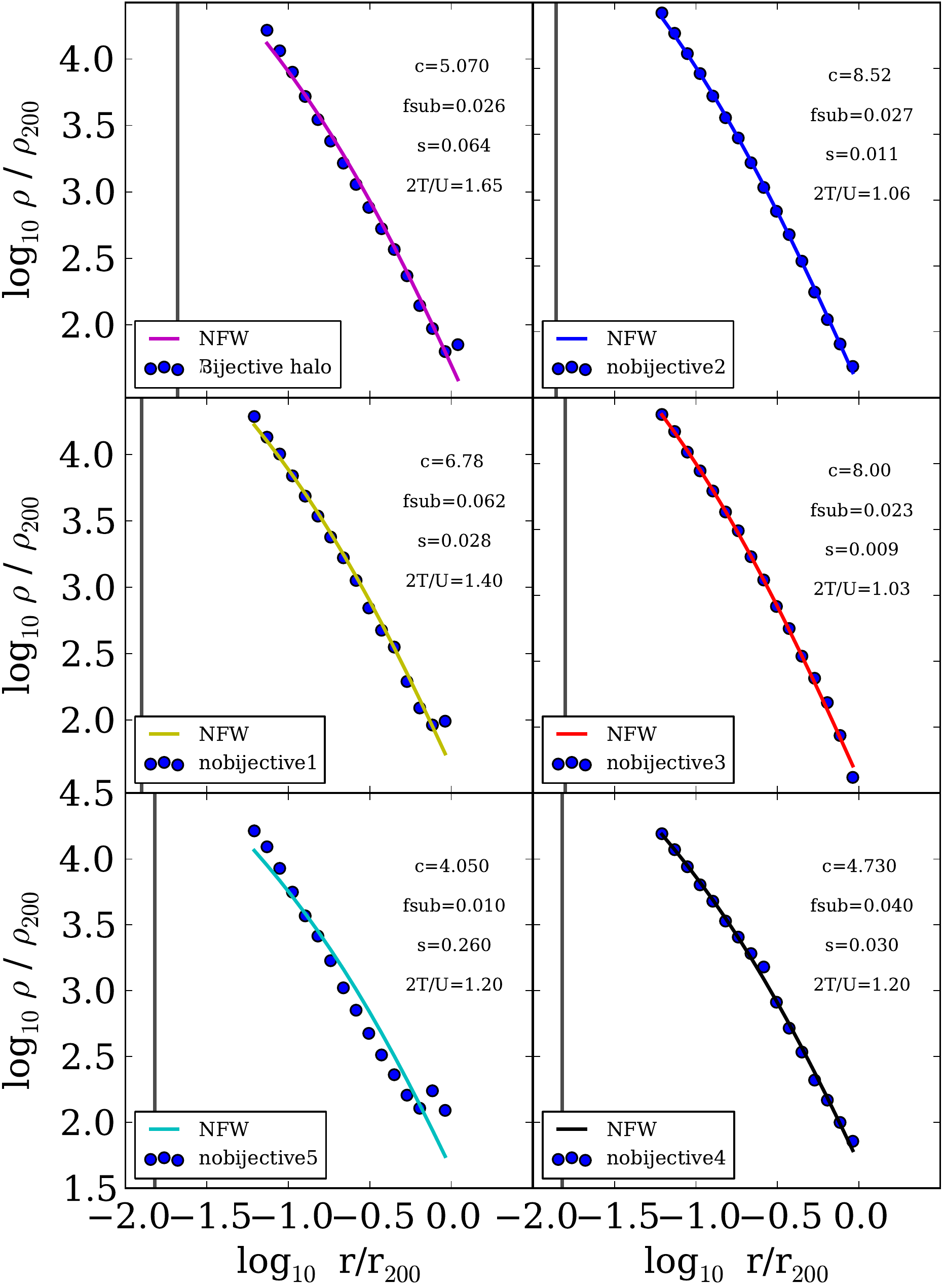}
\caption{Density profiles, $\rho(r)$, for each of the \Dhaloes shown
in Fig.~\ref{fig:nonbi}. The colour of the fitted NFW curve matches the
colour coding of the individual \Dhaloes in Fig.~\ref{fig:nonbi}. The
two-parameter, $\delta_{c}$ and $r_{\rm s}$, NFW least-square fits were
performed over the radial range $0.05 < r/r_{200} <1$, shown by the
black circles in Fig.~\ref{fig:nonbi}.  The minimum fit radius
$r/r_{200} =0.05$ is always larger than the convergence radius derived
by Power et al (2003), which we indicate by the solid vertical line in
each panel.}
\label{fig:den1}
\end{figure}

Fig.~\ref{fig:nonbi} shows a single \FOF halo and its component
\Dhaloes which we use to illustrate the application of these selection
criteria and ability of NFW profiles to fit secondary/non-bijective \Dhaloes.
The spherically averaged density profiles and our NFW fits
to each of these \Dhaloes are shown in Fig.~\ref{fig:den1} along with
the values of the three selection parameters $f_{\rm sub}$, $s$ and $2T/|U|$.
The top left panel of Fig.~\ref{fig:den1} shows the density profile
and NFW fit for the main component of the \FOF halo, which can be
identified by the cyan circle in Fig.~\ref{fig:nonbi} 
which marks twice the half mass radius
the most massive substructure in the \FOF halo. In previous
analyses of \FOF haloes, such as \cite{neto07}, this would be the only
density profile fitted to the mass distribution shown in Fig.~\ref{fig:nonbi}.
The bijectively matched \Dhalo has the same centre as the \FoF halo
and the NFW fit
is performed on all the mass within $r_{200}$, (indicated by the
concentric black circle) consequently 
the density profile and NFW fit of the
bijectively matched \Dhalo is necessarily identical to that or the corresponding \FoF halo.
Examining this
region in Fig.~\ref{fig:nonbi},  we can clearly see that the mass distribution
is asymmetric and has several distinct substructures indicative of a
recent merger. This halo is not
relaxed according to the above selection criteria as it fails to
satisfy the cut on $2T/|U|$. Also its value of the centre offset, $s$,
comes close to the threshold. The NFW fit to its density profile can
be seen to have significant deviations at both large and small radii.

We are also interested in whether NFW profiles provide acceptable fits
to the other \Dhaloes found within this single \FOF halo.  These are
shown in the remaining panels of Fig.~\ref{fig:den1}.  According to
the selection criteria three of these \Dhaloes (those in the
right-hand column) are relaxed. These are the blue, red and black
\Dhaloes in Fig.~\ref{fig:nonbi} and their density profiles are shown,
respectively, in the top, middle and bottom right-hand panels of
Fig.~\ref{fig:den1}.  In all cases we see that the NFW fits provide a
good description of the mass profile of these relaxed \Dhaloes. The
remaining two \Dhaloes fail one or other of the selection
criteria. The yellow \Dhalo of Fig.~\ref{fig:nonbi}, whose density
profile is shown in the middle-left panel of Fig.~\ref{fig:den1},
marginally fails the cut on $2T/|U|$.  The cyan \Dhalo of
Fig.~\ref{fig:nonbi}, whose density profile is shown in the
bottom-left panel of Fig.~\ref{fig:den1}, which strongly exceeds the
threshold on $s$, can be seen to be very poorly fit by the NFW profile
and have a particularly low concentration.  This \Dhalo is very close
to being within twice the half mass radius of the most massive
substructure of the \FOF halo, marked by the cyan circle in
Fig.~\ref{fig:nonbi}. This being the radius used by the \Dhalo
algorithm as part of its criteria to determine whether two \subhaloes
should be considered as two distinct haloes or components of the same
halo. It is this proximity to a merger that both creates the large
offset, $s$, between the potential minimum and the centre of mass
within $r_{200}$ and distorts the object's density profile.  We also
note that this \Dhalo has the most extreme ratio of $r_{200}$ to twice
its half mass radius. In Fig.~\ref{fig:fof} we saw that for isolated
haloes $r_{200}$ and twice the half mass radius were very comparable,
but in contrast we see in Fig.~\ref{fig:nonbi} that the $r_{200}$ of
secondary \Dhaloes can be significantly boosted by the density of the
surrounding environment.

This systematic difference in the ratio of \Dhalo mass to $M_{200}$
for bijective and non-bijective \Dhalos is illustrated in
Fig.~\ref{fig:nonbi_distrb}
which should be contrasted with the right-hand panel of
Fig.~\ref{fig:ms2}. We see that the scatter in the ratio of
$M_{\rm \Dhalo}/M_{200}$ is considerably larger for the non-bijective 
\Dhaloes than it is for bijective \Dhaloes. For bijective \Dhaloes
the 5 to 95\% range of the distribution spans only a 30\% range in
the ratio of $M_{\rm \Dhalo}/M_{200}$, while this is increased to
approximately a factor of two for the non-bijective \Dhaloes. In 
addition the median $M_{\rm \Dhalo}/M_{200}$ ratio is reduced
from $1.2$ for bijective \Dhaloes to $\approx 0.95$ for 
non-bijective \Dhaloes.  These differences are principally caused
by the way the \subfind algorithm \citep{springel01} is effected by the
local environment. \subfind locates the edge of a substructure by
searching for a saddle point in the density distribution. Hence if
the same sub-structure is placed in a denser environment this will
move the saddle point in and reduce the mass that \subfind associates
with the sub-structure \citep[see][for a detailed discussion]{muldrew11}.
As a \Dhalo mass is simply the sum of the
masses of the \subhaloes from which it is composed this in turn reduces 
the mass assigned to the \Dhalo. This systematic dependence of \Dhalo
mass on environment is one of the reasons why instead of directly
using the \Dhalo mass as input to \GALFORM semi-analytic model we
instead force the halo masses in the halo merger trees to increase
monotonically so that they do not artificially  decrease, just prior to 
mergers, due to such environmental effects.

\subsection{The Mass-Concentration relation}

\begin{figure}

\includegraphics[width=8.5cm]{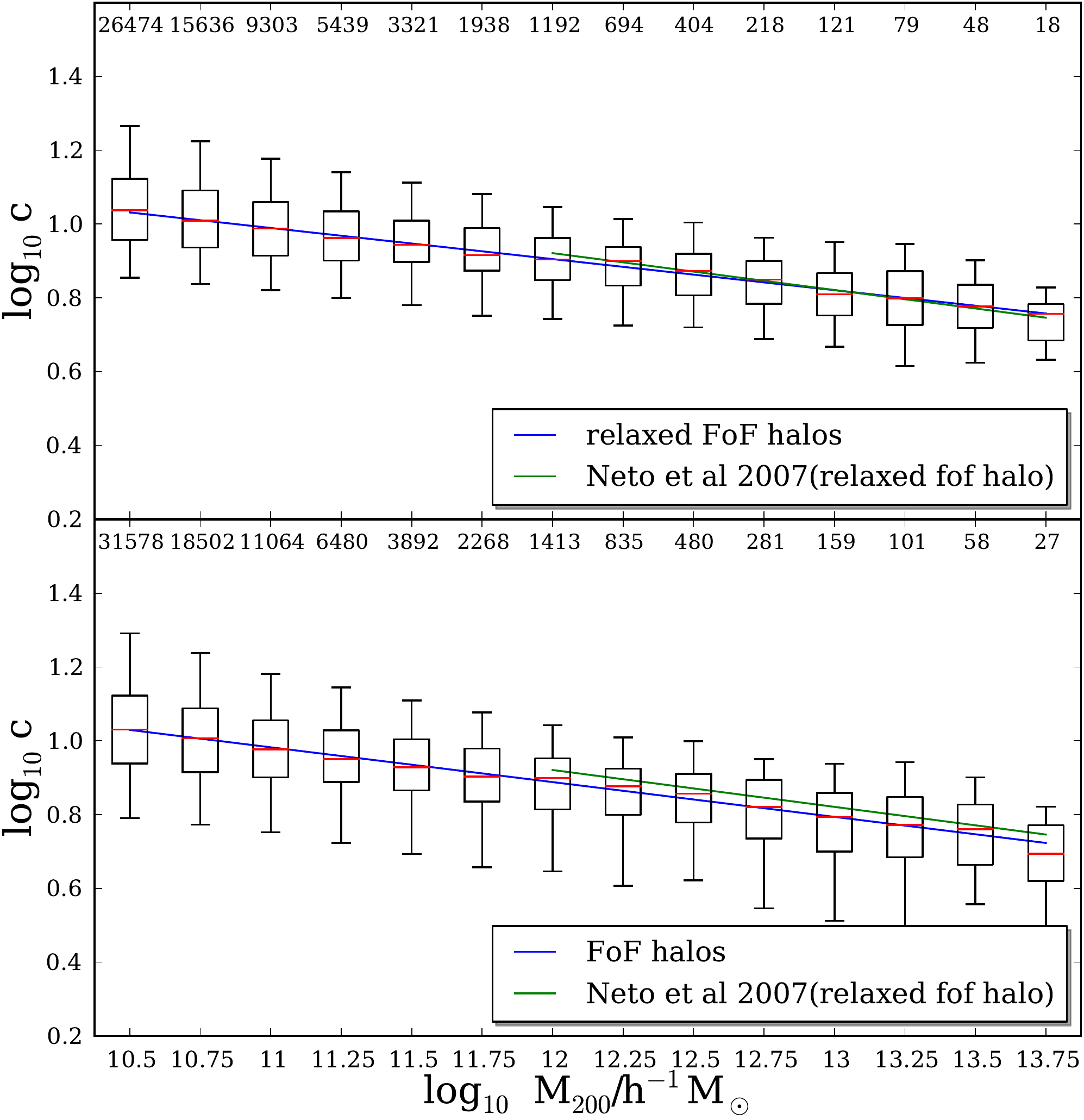}%

\caption{The mass-concentration relation for relaxed FoF haloes
  in MSII (top panel) and for all the FoF haloes (bottom panel). The boxes represent the 25\% and
  75\% centiles of the distribution, while the whiskers show the 5\%
  and 95\% tails. The numbers on the top of each panel indicate the
  number of haloes in each mass bin. The median concentration as a
  function of mass is shown by the blue solid line and is well fit by
  the linear relations given in equations \ref{eq:cmrel} and \ref{eq:cmall}. The green
  lines in each panel correspond to fits of Neto et al (2007).}
 \label{fig:fofmc}
\end{figure}

Here we compare the mass-concentration relation for \FOF haloes that
we find in the high
resolution MSII simulation with that found by \cite{neto07} in the lower
resolution Millennium Simulation.\footnote{As a precise test of our
  methods we first applied our analysis to
  \FoF haloes in the milli-MillenniumII simulation, which has the same
   volume, initial conditions and data format as MillenniumII
  \citep{Boylan-Kolchin09}, 
   but lower mass resolution, equal to that of the Millennium Simulation \citep{springel05b} 
  analysed by \cite{neto07}. We found precise agreement with the 
  mass-concentration relationship published in \cite{neto07}.}
We then go on to compare this relation with the relation we find for the
secondary/non-bijective \Dhaloes. There is no need to separately look at the
bijective \Dhaloes as their $M_{200}$ and $c$ are necessarily the same
as that of the corresponding \FOF haloes as they have the same centre
and all the surrounding mass is used in the fit.
As in \cite{neto07} the mass we use in these
relations is the $M_{200}$ of the NFW fit rather than the directly
measured value.
Fig.~\ref{fig:fofmc} shows concentration as a function of mass for the range
$10^{10.5} < M_{200}/h^{-1} {\rm M}_{\odot} < 10^{13.75}$
for our catalogue of \FoF haloes. The top panel is for our {\it relaxed} \FoF halo sample, while the bottom
panel shows results for all the \FoF haloes, including systems that do
not meet our equilibrium criteria. In each case we find a significant
spread in concentration at fixed mass with a weak trend for decreasing
concentration with increasing mass. This is generally interpreted 
\citep{nfw95,nfw96,nfw97,bullock01,eke01,neto07,gao08} as reflecting the typical
formation time of the halo with the lowest  mass haloes forming
earliest and having high density cores which reflect the density of the
universe at the time they formed.
 The dependence of the median
concentration of \FOF haloes on mass is well described by the power-law fit 
\begin{equation}
c_{200} =  5.45\, \left( 
M_{200}/10^{14} h^{-1} \, {\rm M}_\odot \right)^{-0.084} ,
\label{eq:cmrel}
\end{equation}
for {\it relaxed} haloes and by
\begin{equation}
  c_{200} =  5.01 \, \left( 
M_{200}/10^{14} h^{-1}\, {\rm M}_\odot \right)^{-0.094} 
\label{eq:cmall}
\end{equation}
for all haloes. These fits were performed only over the mass range 
$10^{10.5} < M_{200}/h^{-1} {\rm M}_{\odot} < 10^{13.75}$ due to poor
statistics at higher masses
and are shown by
the blue solid lines in Fig.~\ref{fig:fofmc}.
Also shown on Fig.~\ref{fig:fofmc} is the fit for the median
concentration for relaxed haloes found by \cite{neto07}. 
We plot these green lines only for $M_{200}>10^{12} /h^{-1} M_{\odot}$
corresponding to the resolution limit of their study. We see that over
the overlapping mass range our median concentrations agree very well
with those of \cite{neto07} indicating that the mass profiles over the
fitted radial range, $-2.5<\log(r/r_{200})<0$, are not affected by mass resolution. Our fit is
also similar to  the relation $c_{200} = 5.6(M_{200}/10^{14} h^{-1} M_{\odot})^{-0.098}$
found by \cite{maccio07} for relaxed haloes. The small difference
could be because they fit the mean rather than median of the relation or
due to differences in the criteria used to select relaxed haloes.
Like us and \cite{neto07}, \cite{maccio07} find unrelaxed haloes have
systematically lower concentrations.

Having demonstrated that for \FOF haloes we recover a
mass-concentration relation which is in very accurate agreement with
previous work \citep{neto07,maccio07}, we now want to compare
mass-concentration relations for our bijective and non-bijective
\Dhaloes. The mass-concentration relation we find for the bijective
\Dhaloes is practically identical that of the \FOF haloes plotted in
Fig.~\ref{fig:fofmc} and so we have chosen not to effectively repeat
the same plot. The similarity is inevitable as Fig.~\ref{fig:f1}
shows that for masses greater than $10^{10.5}~h^{-1} M_\odot$,
for which we can measure concentrations, the fraction of \FOF haloes
that have bijective matches with \Dhaloes is greater than 95\% and
these bijectivey matched haloes have identical centres and so
identical fitted NFW mass profiles.

\begin{figure}
\includegraphics[width=8.5cm]{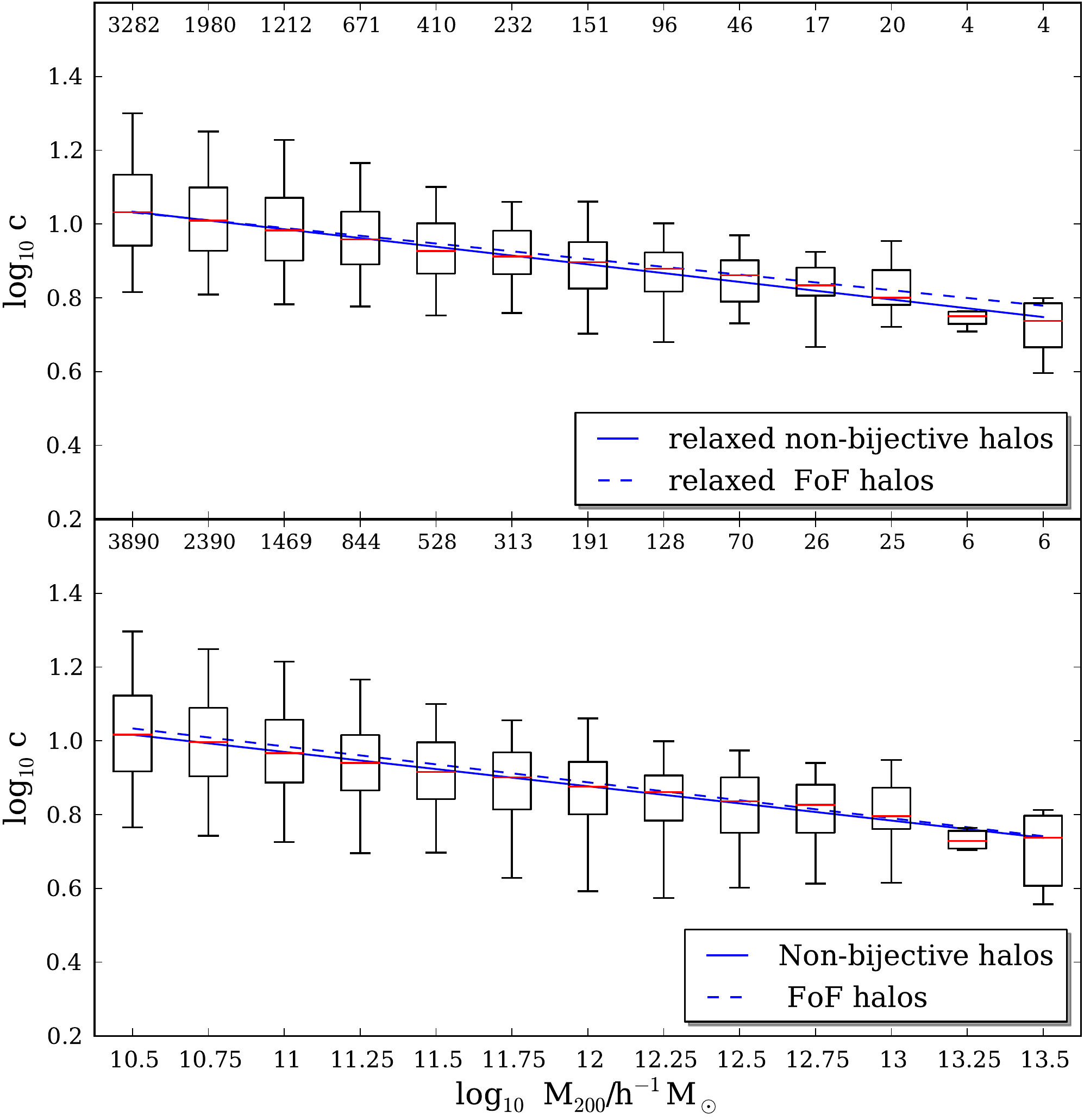}%
\caption{ The mass-concentration relation for relaxed non-bijective
\Dhaloes in MSII (top panel) and for all the non-bijective
\Dhalos (bottom panel). The boxes represent the 25\% and 75\% centiles
of the distribution, while the whiskers show the 5\% and 95\%
tails. The numbers on the top of each panel indicate the number of
haloes in each mass bin. The median concentration as a function of
mass is shown by the blue solid line and is well fit by the linear
relations given in equations \ref{eq:cmrelnb} and \ref{eq:cmallnb}.
The blue dashed line in
each panel repeats the fits to the median mass-concentration relation
for \FoF  haloes shown in Fig.~\ref{fig:fofmc}}
\label{fig:dhalomc}
\end{figure}

In Fig.~\ref{fig:dhalomc} we show the mass-concentration for relaxed
and all non-bijective
\Dhaloes. These haloes are all secondary fragments of \FOF haloes and
so are a completely disjoint catalogue of haloes to those represented
in the \FOF mass-concentration relations of Fig.~\ref{fig:fofmc}.
To aid in comparing the two sets of relations we plot the power-law fits to the
median mass-concentration relations of Fig.~\ref{fig:fofmc} as dashed
lines in Fig.~\ref{fig:dhalomc}. It can be seen that these are very
similar to the power-law fits to the median relations 
\begin{equation}
c_{200} =  4.90\, \left( 
M_{200}/10^{14} h^{-1} \, {\rm M}_\odot \right)^{-0.093} ,
\label{eq:cmrelnb}
\end{equation}
for {\it relaxed} and 
\begin{equation}
  c_{200} =  5.01\, \left( 
M_{200}/10^{14} h^{-1}\, {\rm M}_\odot \right)^{-0.095} 
\label{eq:cmallnb}
\end{equation}
for all the non-bijective \Dhaloes which are shown by the solid lines in
Fig.~\ref{fig:dhalomc}.

\begin{figure*}
\includegraphics[width=15cm]{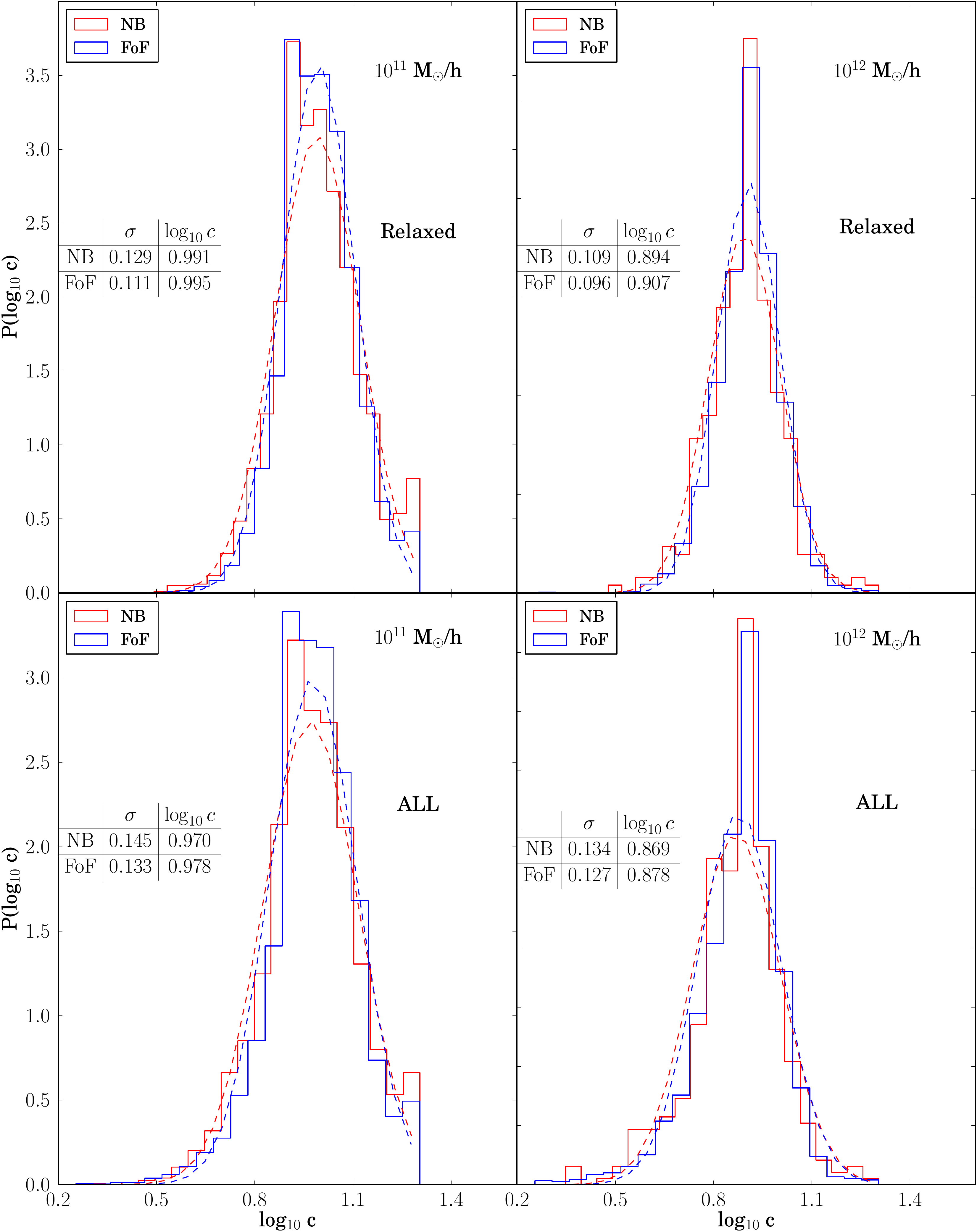}
\caption{The distribution of concentrations for haloes in the two mass
 bins $10.75 < \log_{10} M_{200}/h^{-1}{\rm M}_\odot <11.25$ and
 $11.75 < \log_{10} M_{200}/h^{-1}{\rm M}_\odot <12.25$.  The upper
 panels are for samples of relaxed haloes while the bottom panels are
 for all haloes whether or not they satisfy the relaxation
 criteria. In each panel the blue histogram is for \FOF haloes and the
 red histogram is for \Dhaloes that do not have bijective matches to
 \FOF haloes.  The smooth curves are log-normal approximations with
 the same $\log_{10}c$ and second moment, $\sigma$, as the measured
 distributions. The corresponding values of $\log_{10}c$ and $\sigma$
 are given in the legend. }
\label{fig:his}
\end{figure*}

Comparison of the bars and whiskers in Fig.~\ref{fig:fofmc} and
Fig.~\ref{fig:dhalomc} show that the not only do the median
mass-concentration relations
for \FOF and non-bijective \Dhaloes agree very well, but the
distribution of concentrations about the medians are also quite
similar. The large number of haloes we have in the MII simulation
enables us to look at these distributions in more detail and in 
Fig.~\ref{fig:his} we show histograms of the concentration,
distributions along with log-normal approximations
\begin{equation}
P(\log_{10}c)=\frac{1}{\sqrt{2\pi}\,\sigma}
\exp{\left[-\frac{1}{2}\left(
\frac{\log_{10}c-\left\langle\log_{10}c\right\rangle}{\sigma}
\right)^2 \right]}, 
\end{equation}
for two mass bins centred on $10^{11}$ and
$10^{12}$~\Msol.
We see in all cases that the non-bijective \Dhaloes have a very
similar distribution of concentrations as the distribution of the 
corresponding \FOF sample and that both are 
approximated accurately by log-normal distributions. 
Note that in both cases we are binning haloes by the $M_{200}$ of
their fitted NFW profile and so we are affected by the \Dhalo mass
being perturbed and suppressed in non-bijective \Dhaloes.
We recall that the \FOF
sample is essentially the same as the sample of bijectively matched \Dhaloes and so
we conclude that concentration distribution is essentially the same
for both the primary \Dhaloes and those that are secondary fragments
of \FOF haloes.  In all cases the concentration distributions
for the relaxed samples have slightly higher
median concentrations and smaller dispersions than the corresponding
complete mass selected samples.

Also of interest is the fraction of both \FoF haloes and non-bijective
\Dhaloes that satisfy the equilibrium criteria.  From the number of
objects per mass bin given in the labels on Figs.~\ref{fig:fofmc} and~\ref{fig:dhalomc} 
this can be seen to be in the range of 80 to 85\%
for both \FOF and \Dhaloes.  One might at first expect that many
multi-nucleated \FOF haloes would fail both the threshold on the
asymmetry, $s$, and the fraction of mass in sub-structures, $f_{\rm
sub}$. However as these statistics are evaluated only using the mass
within $r_{200}$ and not across the whole \FOF halo, $\gsim 98$\% of
\FoF haloes pass the substructure threshold and $\gsim 88$\% the
asymmetry threshold. The first of these numbers is slightly lower for
the non-bijective \Dhalos, i.e. only $\gsim 93$\% pass the
substructure threshold. However those passing the more stringent
asymmetry threshold is more comparable at $\gsim 86$\%, while for both
\FOF and non-bijective \Dhalos $\gsim 93$\% pass the criterion that
the virial ratio $2T/|U|<1.35$. Consequently the fraction
of the non-bijective \Dhalos that pass the relaxation criteria is very similar to that for
 the \FOF or bijective \Dhalos. Hence in both cases the mass-concentration
distributions that we have quantified are representative of the vast
majority of the haloes.

\section{Conclusions}
\label{sec:conc}

We have used the high resolution Millennium Simulation II cosmological
N-body simulation to quantify the properties of haloes defined by the
\Dhalo algorithm. This algorithm is designed to produce merger trees
suitable for use with the semi-analytic galaxy formation model,
\GALFORM.  We have included a full description of the \Dhalo algorithm
which produces a set of haloes, and the merger trees that describe
their hierarchical evolution, that are consistent between subsequent
snapshots of the N-body simulations.  We have presented the properties
of the \Dhaloes by comparing them with the corresponding properties of
the much more commonly used \FoF haloes \citep{davis85}.

We have shown that unlike the \FOF algorithm the \Dhalo algorithm is
successful in avoiding distinct mass concentrations being prematurely
linked together into a single halo when their diffuse outer haloes
touch. We have also illustrated how some \Dhaloes can be composed of
more than one \FOF halo. This occurs as structure formation in CDM
models is not strictly hierarchical and occasionally a halo, after
falling into a more massive halo, may escape to beyond the virial
radius of the more massive halo. For the purposes of the \GALFORM
semi-analytic model it is convenient to consider such haloes as
remaining as satellites of the main halo. We find that such remerged
\FOF haloes are not uncommon, but contribute very little mass to the
larger haloes to which they are (re)attached.

Despite the complex mapping between \FOF and \Dhaloes, which results
in a significant fraction of \FOF haloes being broken up into multiple
\Dhaloes while other \FOF haloes get (re)merged into a single \Dhalo, we
find that the overall mass functions of the two sets of haloes are
very similar. The mass functions of our 
\Dhalo and \FOF halo catalogues are both reasonably well fit over the mass
range of $10^{8}$ to $10^{13.5}$~\Msol  by currently
popular analytic mass functions such as those of Warren et al (2006) and
Reed et al. (2007).

Approximately 90\% of the \Dhaloes have a unique one-to-one,
bijective, match with a corresponding \FOF halo. For this subset of
haloes the mass of the \Dhalo, $M_{\rm \Dhalo}$, correlates much more
closely with the standard virial mass, $M_{200}$, than does the \FOF
mass.  The median $M_{\rm \FOF}/M_{200}=1.2$ and 90\% of the distribution of
this mass ratio spans a factor $1.9$, while for the same \Dhaloes
the median $M_{\rm \Dhalo}/M_{200}=1.15$ and corresponding width of the
distribution spans only a factor $1.3$. The larger scatter in the
\FOF case is often caused by secondary mass concentrations that lie
outside the $r_{200}$ radius of the main substructure and are linked into
the \FOF halo by particle bridges in overlapping diffuse haloes.  The
non-bijective \Dhaloes have a wider distribution, with 90\% of the
distribution spanning a factor 2.2 and with the median ratio reduced to
$M_{\rm \Dhalo}/M_{200}=0.95$. This is due to the \SUBFIND substructure
finder, which is part of the \Dhalo algorithm, assigning less mass to
\subhaloes when they move into overdense environments.  When utilised
in \GALFORM this systematic loss of mass is not an issue as the
merger trees are preprocessed and mass is added back in to ensure
the branches of the \GALFORM merger trees always have monotonically
increasing masses.

The high resolution of the Millennium~II simulation has allowed us to
study the density profiles and concentrations of both \FOF and
\Dhaloes over a wide range of mass. To avoid contaminating our samples
with unrelaxed haloes for which fitting smooth spherically symmetric
profiles is inappropriate we exclude unrelaxed haloes using the
relaxation criteria from Neto et al (2007).  We find that 80\% of both
\FOF and \Dhaloes are relaxed according to these criteria.  For \FOF
haloes we accurately reproduce the mass--concentration distribution
found by Neto et al (2007) at high masses and extend the distribution
to much lower masses.  Combining our results with those of
~\citet{maccio07} and ~\citet{neto07}, we find that a single power law
reproduces the mass-concentration relation for over five decades in
mass. We also find that the mass-concentration distributions for
\Dhaloes agree very accurately with those for \FOF haloes. This is
true even for non-bijective \Dhaloes which are secondary
components of \FOF haloes.  The properties of such haloes have
generally been overlooked in previous studies.
We show that the distributions of concentrations around the
mean  mass-concentration relation are well described
by log-normal distributions for both the \FOF and \Dhaloes.

\section*{acknowledgements}

This work was supported by the Science and Technology Facilities
[grant number ST/F001166/1].  LJ acknowledges the support of a Durham
Doctoral Studentship.  This work used the DiRAC Data Centric system at
Durham University, operated by the Institute for Computational
Cosmology on behalf of the STFC DiRAC HPC Facility
(www.dirac.ac.uk). This equipment was funded by BIS National
E-infrastructure capital grant ST/K00042X/1, STFC capital grant
ST/H008519/1, and STFC DiRAC Operations grant ST/K003267/1 and Durham
University. DiRAC is part of the National E-Infrastructure.

\appendix
\section{Constructing \Dhalo Merger Trees}
\label{app:Trees}

Here we describe in detail the algorithm used to produce the Dhalo
merger trees. These merger trees are intended to be used as input to
the \galform semi-analytic model of galaxy formation. The need for
consistency between the halo model used in the semi-analytic
calculation and the N-body simulation imposes some requirements on the
construction of the merger trees.

The \galform galaxy formation model makes the approximation that
mergers between haloes are instantaneous events and assumes that
haloes, once merged, do not fragment. However, in N-body simulations
halo mergers take a finite amount of time and it is not uncommon for a
halo falling into another, more massive halo to escape to well beyond
the virial radius after its initial infall
\citep{gill05,ludlow09}. We therefore need to choose when to
consider N-body haloes to have merged in the semi-analytic model and
define our haloes such that they remain merged at all later times. We
also wish to define the haloes used to construct the trees such that,
as far as possible, they resemble the spherically symmetric,
virialised objects assumed in the galaxy formation model. Quantifying
the extent to which we have achieved this is one of the main aims of
this paper.

\subsection{Halo catalogues}
\label{sec:groups}

The first step in building the merger trees is to use the \FoF
\citep{davis85} and \subfind algorithms
\citep{springel01} to identify haloes and \subhaloes in all of
the simulation snapshots. The \subfind algorithm decomposes each \FoF
halo into \subhaloes by identifying self bound density maxima. 
Usually the most massive \subhalo contains most of the mass of the
original \FoF halo. Secondary density maxima give rise to additional
\subhaloes. Compared to the \FoF halo the most massive \subhalo
does not include any of the mass assigned to other \subhaloes 
(a simulation particle can only belong to one \subhalo) nor does it
include particles that are not gravitationally bound to it.
Some of the lowest mass \FoF haloes have no self
bound \subhaloes and most \FoF haloes have at least some ``fuzz''
of unbound particles which belong to no \subhalo. 
\FoF haloes with no self-bound
\subhaloes are not used in the construction of the merger trees.

\subsection{Building the \subhalo merger trees}
\label{app:build}

Before we can construct the \Dhalo merger trees, it is necessary to
define \subgroup merger trees by identifying the descendant of each
\subgroup. The code we use to do this was included in the merger trees
comparison project carried out by \cite{srisawat13} under the
name {\sc D-Trees}. The project concluded that it was desirable feature
for a merger tree
code to use particle IDs to match haloes between snapshots and have
the ability to search multiple snapshots for descendants. The latter
requirement was due to the tendency of the AHF group finder
\citep{kk09} used in the project to temporarily fail to
detect sub-structures during mergers.

Since \subfind suffers from a similar problem, we allow for the
possibility that the descendant of a \subgroup may be found more than
one snapshot later. Our approach is to devise an algorithm which can
identify the descendant of a halo at any single, later snapshot,
apply it to the next $N_{\rm step}$ snapshots (where $N_{\rm step}
=5$), and pick one of these $N_{\rm step}$ possible descendants
to use as the descendant of the \subgroup in the merger trees.

Alternative solutions to this problem include allowing the merger tree
code to modify the \subgroup catalogue to ensure consistency of
\subgroup properties between snapshots ({\sc ConsistentTrees},
\citealt{behroozi13}) and using information from previous
snapshots to define the \subgroup catalogue (HBT,
\citealt{han12}).

In common with all but one of the merger tree codes in the comparison
({\sc Jmerge}, which relies entirely on aggregate properties of the
haloes), we identify descendants by finding \subgroups at different
snapshots which have particles in common.

\begin{figure*}
\includegraphics[width=\textwidth]{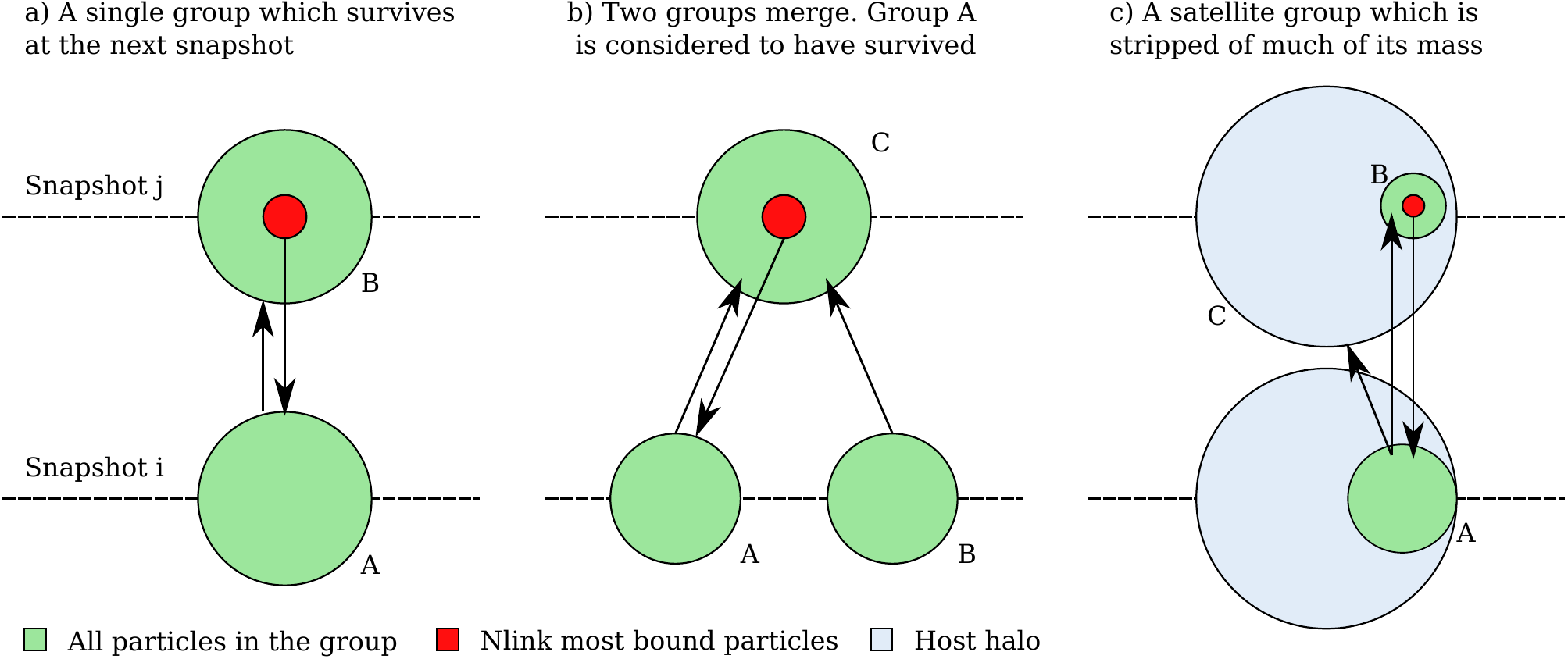}
\caption{Schematic examples illustrating the method used to link
  \subfind \subhaloes between pairs of snapshots $i$ and $j$, where $i<j$.
  The green circles represent \subfind \subhaloes. The most bound
  $N_{\rm link}$ particles in each \subhalo at the later time are shown in
  red. From left to right are a) a single, isolated \subhalo which still
  exists at the next snapshot, b) a merger between \subhaloes \emph{A} and
  \emph{B} where more of the most bound partciles of the merged halo
  \emph{C} come from halo \emph{A} than from any other halo and
  therefore halo \emph{A} is considered to be the main progenitor of
  halo \emph{C}, and c) a satellite \subgroup orbiting within a
  background halo which loses a large fraction of its particles to its
  host halo at the next snapshot but is still identified by
  \subfind. Arrows between green circles show the location of the
  majority of the particles in the \subhalo at the later snapshot. Arrows
  starting from red circles show the location of the majority of the
  most bound particles at the earlier snapshot.}
\label{fig:link_snaps}
\end{figure*}

\subsubsection{Identifying a descendant at a single, later snapshot}
\label{app:findingdescendants}

To find the descendant at snapshot $j$, of a halo which exists at an
earlier snapshot, $i$, the following method is used. For each halo
containing $N_p$ particles the $N_{\rm link}$ most bound are
identified, where $N_{\rm link}$ is given by
\begin{equation}
N_{\rm link} = \min(N_{\rm linkmax}, \max(f_{\rm trace} N_{p}, N_{\rm linkmin}))
\end{equation}
with $N_{\rm linkmin} = 10$, $N_{\rm linkmax} = 100$ and $f_{\rm trace}=0.1$.

For each of the haloes at snapshot $i$, descendant candidates are
found by locating all haloes at snapshot $j$ which 
received at least one particle from the earlier halo. Then, a single
descendant is chosen from these candidates as follows. If any of the
descendant candidates received a larger fraction of their $N_{\rm link}$
most bound particles from the progenitor halo than from any other
halo at the earlier snapshot, then the descendant is chosen from
these candidates only and the halo at snapshot $i$ will be designated
the main progenitor of the chosen descendant; otherwise, all
candidates are considered and the halo will not be the main
progenitor of its descendant. The descendant of the halo at snapshot
$i$ is taken to be the remaining candidate which received the largest
fraction of the $N_{l\rm ink}$ most bound of the progenitor
halo. For each halo at snapshot $j$, this method identifies zero or
more progenitors of which at most one may be a main progenitor. Note
that it is not guaranteed that a main progenitor will be found for
every halo.

By following the most bound part of the \subgroup, we ensure that if
the core of a \subgroup survives at the later snapshot it is identified
as the descendant irrespective of how much mass has been lost. It also
means that in cases where an object at the later snapshot has multiple
progenitors we can determine which one of the progenitors contributed
the largest fraction of the most bound core of the descendant
object. We consider this main progenitor to have survived the merger
while the other progenitors have merged onto it and ceased to exist as
independent objects.

Fig.~\ref{fig:link_snaps} shows three examples of this linking
procedure. In the simplest case (left) a single, isolated \subgroup
\emph{B} at snapshot $j$ is identified as the descendant of \subgroup
\emph{A} which exists at the earlier snapshot $i$. Since more of the
most bound particles of \subgroup \emph{B} come from \subgroup \emph{A}
than from any other \subgroup, we conclude that \emph{A} is the main
progenitor of \emph{B}. In the second case (centre) two \subgroups
\emph{A} and \emph{B} merge to form \subgroup \emph{C} at the later
snapshot. \Subgroup \emph{A} is determined to be the main progenitor
because it contributed the largest fraction of the most bound
particles of the descendant, \emph{C}. In the third example (right) a
satellite \subgroup \emph{A} exists within a more massive host halo. In
this case, particles from the \subgroup \emph{A} are split between
\subgroup \emph{B} and the host halo \emph{C} at the later
snapshot. While a large fraction (or even the vast majority) of the
particles from \subgroup \emph{A} may belong to the host halo at the
later snapshot, we choose \subgroup \emph{B} as the descendant because
its most bound part came from \subgroup \emph{A}.

\begin{figure}
\begin{center}
\includegraphics[width=0.8\linewidth]{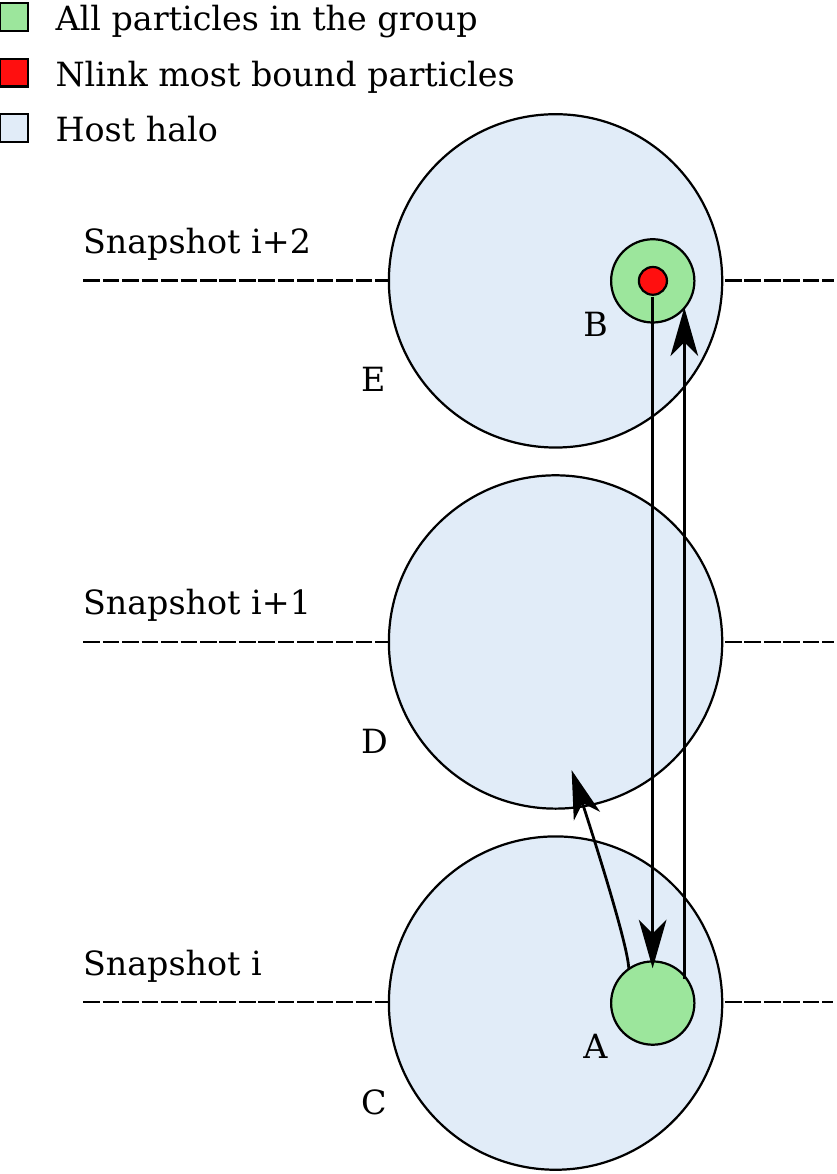}
\end{center}
\caption{A schematic example of a case where the descendant of a
  \subgroup is found to be more than one snapshot later. The green
  circles represent a satellite \subfind \subgroup within a larger host
  halo which is represented by the blue circles. Three consecutive
  snapshots are shown.}
\label{fig:skip_snapshot}
\end{figure}

\subsubsection{Searching multiple snapshots for descendants}
\label{app:descendantsnapshots}

If a \subgroup is not found to be the main progenitor of its descendant,
this may indicate that the \subgroup has merged with another \subgroup and no
longer exists as an independent object. However, it is also possible
that the substructure finder
has simply failed to identify the object at the
later snapshot because it is superimposed on the dense
central parts of a larger \subhalo. 
Typically this phase lasts for a small
fraction of the host halo dynamical time \citep{behroozi13}
which in turn is much shorter
than the usual interval between the snapshots of cosmological N-body
simulations. Hence by looking one snapshot ahead we will normally find
the missed \subhalo, but one can be unlucky and catch it half an orbit
later when again it is hidden by the dense core of the more massive
subhalo in which it is orbiting. Hence looking several snapshots
ahead exponentially suppreses this possibility.
Thus in order to distinguish between \subgroup mergers and \subgroups
which are just temporarily lost it is
necessary to search multiple snapshots for descendants.

In our algorithm for each snapshot 
$i$ in the simulation descendants are identified at later
snapshots in the range $i+1$ to $i+N_{\rm step}$ using the method
described in section~\ref{app:findingdescendants}. For each \subgroup at
snapshot $i$ this gives up to $N_{\rm step}$ possible descendants. One of
these descendants is picked for use in the merger trees as follows: if
the \subgroup at snapshot $i$ is the main progenitor of one or more of the
descendants, the earliest of these descendants which does not have a
main progenitor at a snapshot later than $i$ is chosen. If no such
descendant exists, the earliest descendant found is chosen
irrespective of main progenitor status.
 
Descendants more than one snapshot later are only chosen in cases
where the earlier \subgroup is the main progenitor --- i.e. where the
group still survives as an independent object. If the \subgroup does not
survive we have no way to determine whether it merged immediately or
if \subfind failed to detect it for one more snapshots prior to the
merger, so we simply assume that the merger happened between snapshots
$i$ and $i+1$.

Fig.~\ref{fig:skip_snapshot} shows a case where a descendant more than
one snapshot later is chosen. \Subgroup \emph{A} exists at snapshot
$i$. Its descendant at snapshot $i+1$ is found to be the \subgroup
\emph{D}. However, the most bound particles of \emph{D} were not
contributed by \subgroup \emph{A}, but by another progenitor, \subgroup
\emph{C}. This means that \emph{A} is not the main progenitor of its
descendant at snapshot $i+1$ and so it is necessary to consider
possible descendants at later snapshots. Two \subgroups at snapshot
$i+2$ (\emph{B} and \emph{E}) receive particles from \subgroup
\emph{A}. Since the most bound particles of \subhalo \emph{B} came from
\subhalo \emph{A}, \emph{A} is the main progenitor of \emph{B} and
\subgroup \emph{B} is taken to be the descendant of \emph{A}.

\subsection{Constructing a halo catalogue}
\label{app:Dhalo}

\begin{figure}
\begin{center}
\includegraphics[width=0.8\linewidth]{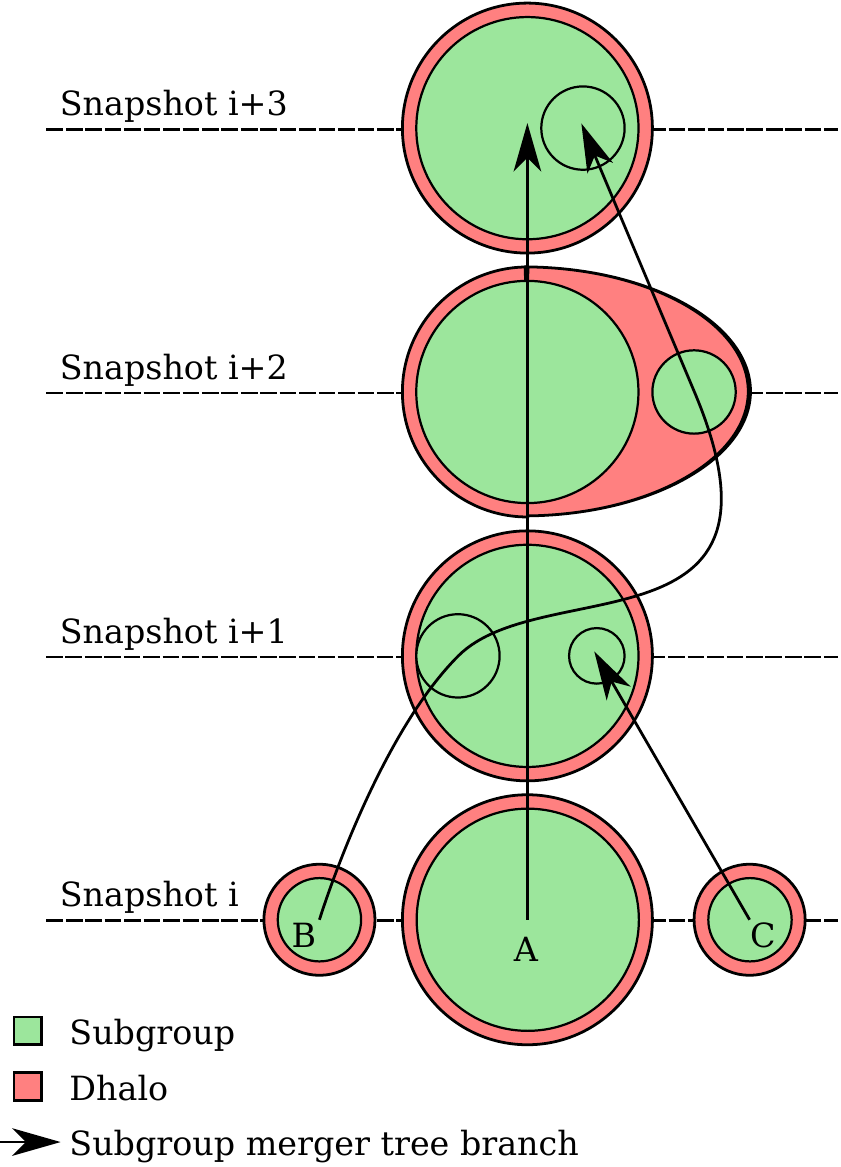}
\end{center}
\caption{An example of a Dhalo merger tree showing two less massive
  haloes falling into another, more massive halo. \Subgroups are shown
  in green. Red areas indicate \subgroups which belong to the same
  Dhalo. The black arrows show branches of the \subgroup merger tree.}
\label{fig:merger_tree}
\end{figure}

At this point we have a descendant for each \subgroup. This is
sufficient to define merger trees for the \subgroups. These \subfind
trees can be split into ``branches'' as follows. A new branch begins
whenever a new \subgroup forms (i.e. the \subgroup has no progenitors). The
remaining \subgroups that make up the branch are found by following the
descendant pointers until either a \subgroup is reached that is not the
main progenitor of its descendant, a \subgroup is reached that has no
descendant, or the final snapshot of the simulation is reached. Each
of these branches represents the life-time of an independent halo or
sub-halo in the simulation. We construct haloes and halo merger trees
by grouping together these branches of the \subgroup merger trees using
methods which will be described below. We refer to the resulting
collections of \subgroups as ``Dhaloes''. Fig.~\ref{fig:merger_tree} is
an example of a Dhalo merger tree with the \subgroup merger tree
branches marked. In this case there are three branches. Branch
\emph{A} is a single, massive halo which exists as an independent halo
at all four snapshots. Branch \emph{B} is a smaller halo which becomes
a satellite \subhalo within halo \emph{A}, but continues to
exist. Branch \emph{C} is another small halo which briefly becomes a
satellite before merging with \emph{A}.

For each \subgroup in a \FoF halo we identify the least massive, more
massive ``enclosing'' \subgroup in the same \FoF halo. \Subgroup \emph{A} is
said to enclose \subgroup \emph{B} if \emph{B}'s centre lies within twice the half
mass radius of \emph{A}. A pointer to the enclosing \subgroup 
is stored for each
\subgroup that is enclosed. This produces a tree structure which is
intended to represent the hierarchy of haloes, sub-haloes, sub-sub-haloes
etc. in the \FoF halo. Any \subgroup which is not enclosed by any
other becomes a new \Dhalo. Any \subgroups enclosed by this \subgroup are
assigned to the new \Dhalo.

We then iterate through the snapshots from high redshift to low
redshift. For each \subgroup we find the maximum number of
particles it ever contained while it was the most massive \subgroup in its
parent \FoF halo. If a satellite \subgroup in a \Dhalo retains a
fraction $f_{\rm split}$ of its maximum isolated mass then it is split
from its parent \Dhalo and becomes a new \Dhalo. Any \subgroups
enclosed by this \subgroup are assigned to the new \Dhalo too. We
usually set $f_{\rm split} = 0.75$, so that when a halo falls into
another, more massive halo the two haloes will only be considered to
have merged into one once the smaller halo has been stripped of some
of its mass. This is to ensure that haloes artificially linked by the
\FoF algorithm are still treated as separate objects.

In some cases a \subgroup may escape from its parent halo. This happens
to halo \emph{B} in Fig.~\ref{fig:merger_tree}. For the purposes of
semi-analytic galaxy formation modelling, we would like to continue to
treat such \subgroups as satellites in the parent halo so that each
in-falling halo contributes a single branch to the halo merger
tree. This is done by merging such objects back on to the Dhalo they
escaped from; the \subgroup is recorded as a satellite within the
original Dhalo at all later times regardless of its spatial
position. Any \subgroups it encloses will also be considered to be part
of this Dhalo.

In practice the re-merging is carried out in the following way. For
each Dhalo \emph{A} we identify a descendant Dhalo \emph{B} by
determining which later Dhalo contains the descendant of the most
massive \subgroup in \emph{A} which survives at the next snapshot. In
every case where a \subgroup in \emph{A} survives, we assign the
descendant of the \subgroup to Dhalo \emph{B}. We repeat this process
for all \Dhaloes at each snapshot in decreasing order of redshift. This
ensures that if any two \subgroups are in the same Dhalo at one
snapshot, and both survive at the next snapshot, they will both be in
the same Dhalo at the next snapshot.

This process produces a \Dhalo catalogue for each snapshot. Each \Dhalo
contains one or more \subgroups and each \subgroup may have a pointer to
a descendant at some later snapshot. Any \subgroups in a \Dhalo which
survive at the next snapshot are guaranteed to belong to the same
Dhalo at the next snapshot. This provides a simple way to identify a
descendant for each \Dhalo and defines the Dhalo merger
trees. Fig.~\ref{fig:merger_tree} shows an example of a \Dhalo merger
tree. The two smaller haloes \emph{B} and \emph{C} merge with a
larger halo \emph{A}. Halo \emph{C} survives as a satellite for one
snapshot before merging with the descendant of \emph{A}. Halo \emph{B}
also becomes a satellite sub-halo and then temporarily escapes from
the parent halo before falling back in. At all times after the initial
infall it is considered to be part of the parent \Dhalo.

\label{lastpage}

\end{document}